\documentclass[structabstract]{aa}  

\usepackage{graphicx}
\usepackage{color}
\usepackage{epsfig}
\usepackage{supertabular}
\usepackage{longtable,lscape}
\usepackage{subfigure}
\usepackage[varg]{txfonts}
\usepackage{natbib}
\bibpunct{(}{)}{;}{a}{}{,}  
\usepackage{rotating}

\def\kms {\rm{km~s^{-1}}}
\def\apj {ApJ}
\def\apjl {ApJL}
\def\apjs {ApJS}
\def\aj {AJ}
\def\aap {A\&A}
\def\mnras {MNRAS}
\def\arcsec{''}

\def\Mpc {\rm Mpc}
\begin{document}
\title{Comparing galaxy populations in compact and loose groups of galaxies}
\author{Valeria Coenda, Hern\'an Muriel \&  H\'ector J. Mart\'\i nez}
\institute{Instituto de Astronom\'ia Te\'orica y Experimental (IATE), CONICET-Observatorio Astron\'omico,
Universidad Nacional de C\'ordoba. Laprida 854, C\'ordoba, X5000BGR. Argentina.\\
\email{vcoenda;hernan;julian@oac.uncor.edu}}
\date{Received XXX, XXXX; accepted XXX , XXXX }
\abstract
   {}
{We perform a comparison of the properties of galaxies in compact groups, loose groups and in the field
to deepen our understanding of the physical mechanisms acting upon galaxy evolution in different environments.}
{We select samples of galaxies in compact groups identified by McConnachie et al., loose groups identified
by Zandivarez \& Mart\'inez, and field galaxies from the Sloan Digital Sky Survey. We compare 
properties of the galaxy populations in these different environments: absolute magnitude, colour, 
size, surface brightness, stellar mass and concentration. We also study the fraction of red and
early type galaxies, the luminosity function, the colour-luminosity and luminosity-size relations.
}
{The population of galaxies in compact groups differ from that of loose groups and the field.
The fraction of read and early type galaxies is higher in compact groups.
On average, galaxies in compact groups are systematically smaller, more concentrated and
have higher surface brightness than galaxies in the field and in loose groups. 
For fixed absolute magnitude, or fixed surface brightness, galaxies in compact groups are smaller.
}
{The physical mechanisms that transform galaxies into earlier types could be more effective
within compact groups given the high densities and low velocity dispersion that characterise
that particular environment, this could explain the large fraction of red and early type 
galaxies we found in compact groups. Galaxies inhabiting compact groups have undergone a major
transformation compared to galaxies that inhabit loose groups.
}
\authorrunning{Coenda, Muriel \& Mart\'inez}
\maketitle
\section{Introduction} 
\label{sec:intro}
Galaxies inhabit a wide range of environments, from isolated galaxies to the core of 
galaxy clusters and compact groups (hereafter CGs).
There is clear evidence that both the properties of galaxies and the relative fraction of
different type of galaxies depend on the environment (e.g. 
\citealt{Oemler:1974,Dressler:1980,Goto:2003,Blanton:2005}). In low density 
environments, galaxies tend to be blue, star forming and late-type, while dense environments are 
dominated by red, early-type galaxies. However, density is not the only relevant parameter when
characterising the relationship between environment and galaxy properties, moreover, it is
well known that galaxy properties strongly depend on stellar mass (e.g. Kauffmann et al. 2004).
The dynamics of the system, e.g. the relative speed with which galaxies move, or the characteristics 
of the intergalactic medium, e.g. the presence of hot gas, may influence the evolution of galaxies 
and thus modify their properties. Within this scenario, different physical 
processes that can affect the evolutionary history of galaxies have been proposed. 
Galaxy-galaxy interactions, mergers or galaxy harassment can substantially change the structure of 
galaxies and even cause significant loss of mass (e.g. \citealt{TT:1972,Moore:1998}). 
On the other hand, the presence of gas in the intergalactic medium can substantially affect 
galaxies through mechanisms such as ram pressure (e.g. \citealt{GG:1972,Abadi:1999}) 
or strangulation (e.g. \citealt{Larson:1980,Balogh:2000,Kawata:2008}). 
The relative influence of these processes depend on several physical parameters which vary
from one environment to another, therefore comparative studies 
involving different galaxy environments are useful for a more complete understanding
of their effect on galaxy evolution.

Among the various environments, CGs are an extreme case.
Although their densities are among the highest observed, both, the number of members, and the 
velocity dispersion of galaxies are lower than those seen in massive loose groups or 
clusters of galaxies \citep{Hickson:1992}. On the other hand, the number of galaxies 
and the velocity dispersion of CGs and low-mass loose groups may be comparable, although the 
crossing times are substantially different. These similarities and differences between loose groups
(hereafter LGs) and CGs represent an useful scenario to test the influence that 
different physical processes have on the galaxy evolution. 
Moreover, \citet{Diaferio:1994} suggested a tight connection between loose and compact groups. They
estimated that the mean lifetime of a compact configuration is $\sim 1 {\rm Gyr}$ and suggested that 
on this time scale, members may merge and other galaxies in the loose group may join the 
compact configuration.
While the properties of galaxies in CGs and LGs have been extensively studied separately, 
no systematic comparison using homogeneous and statistically significant samples have been performed
so far. \citet{Lee:2004} compared the properties of field and CGs galaxies and found that 
the colours of CGs galaxies differ from those of field galaxies in the 
sense that CGs have a higher fraction of elliptical galaxies. 
\citet{Deng:2008} compared the properties of CGs, isolated, and field galaxies 
and found that, in dense regions, galaxies have preferentially greater concentration index 
and early-type morphology. There are numerous studies on the star formation in CGs: 
\citet{Walker:2010} suggest that the compact group 
environment accelerate the evolution of galaxies from star-forming to quiescent; \citet{Bitsakis:2010} 
found a connection between dynamical state and the star formation rate (SFR) in 
the sense that old CGs host late-type galaxies with slightly lower specific star formation rate than 
in dynamically young groups; \citet{Johnson:2007} also found a connection between the star formation 
and the global properties of groups. \citet{Tzanavaris:2010} estimated the SFR using both 
ultraviolet and infrared information and found that the compact groups environment accelerate the 
galaxy evolution by enhancing the star formation processes and favouring a fast transition to quiescence.

The intra-group medium (IGM) in CGs has been extensively studied and can provide useful information on 
the evolution of galaxies. \citet{Torres:2009} searched for young objects in the intra-group medium of 
several CGs and found that groups are in different stages of interaction. Many CGs show X-ray emission 
associated to the hot intra group medium (HIGM), \citet{Rasmussen:2008} study the 
influence of the HIGM on the galaxy evolution and found that galaxy-HIGM interactions would not be 
the dominant mechanism driving cold gas out of the group members, been tidal interactions the 
most likely process to remove gas from galaxies in CGs.

Historically, the identification of CGs has been performed in projection, which causes 
the detection of spurious systems. While this can be solved with redshift determinations, 
the observed compact configuration could be the result of projection effects within LGs. 
These effects have been quantified by \citet{McConnachie:2008}, \citet{Diaz:2010} and 
\citet{Mendel:2011}.
These studies showed the presence of contamination by chance association of galaxies 
in CGs. \citet{Brasseur:2009} found that only ~30 per cent of the simulated groups are truly compact 
in three dimensions. This contamination effect will bias any observational study. 
Even more radical is the scenario proposed by \citet{Tovmassian:2006} who claim that 
CGs are not different from LGs. However, the general consensus is that a significant fraction
of the compact groups with accordant-velocity members are physically dense systems.
\citet{Hickson:1982} compiled a sample of 100 CGs based on the photographic 
plates of the Palomar Observatory Sky Survey. 
This sample has been extensively studied, however the low number of systems does not allow 
the implementation of statistical studies which could disentangle some
dependencies between galaxy properties and environment. Based on surveys of galaxies, 
other samples of CGs have been identified (e.g. \citealt{Iovino:2003,Lee:2004}). 
Using the original selection criteria of 
\citet{Hickson:1982} and based on the Sixth Data Release of the Sloan Digital Sky Survey 
(DR6, \citealt{dr6}), \citet{McConnachie:2009} 
identified two samples of CGs, one of 2,297 compact groups down to a limiting magnitude of 
$r=18$, and a deeper sample of 74,791 down to $r=21$. 

LGs have been extensively identified 
in large redshifts surveys such as the 2dFGRS \citep{Colless:2001} and SDSS \citep{York:2000}, 
allowing the identification of thousands of loose groups in the nearby universe 
(e.g. \citealt{Merchan:2002,Merchan&Zandivarez:2005,Eke:2004,Yang:2007}).
The dependence of galaxy properties on the group properties has been studied by several authors
(e.g. \citealt{Martinez:2002,Weinmann:2006,MM2:2006,Gerke:2007,Hansen:2009,McGee:2011,Wetzel:2011}).
\citet{Weinmann:2009} investigated 
sizes, concentrations, colour gradients and surface brightness profiles of central and satellite 
galaxies in LG groups. These authors found that at fixed stellar mass, late-type satellite 
galaxies have smaller radii, larger concentrations, lower surface brightness and redder 
colours than late-type central galaxies. This effect is 
not found for the early-type galaxies. Similar results are found by \citet{Maltby:2010}. 
\citet{Nair:2010} found no environmental dependence in the size-luminosity 
relation for early-type galaxies. On the hand, several authors found a clear dependence
of the size-luminosity relationship of early-type galaxies between central and satellite or 
field galaxies (e.g. \citealt{Coenda:2009,Bernardi:2009}).

The purpose of this paper is the comparison between the properties of galaxies in compact 
groups and loose groups using homogeneous samples of galaxies and groups having similar 
spatial distributions.

This paper is organised as follows: in section \ref{sec:sample} we describe the samples
groups and galaxies we use in this paper; in section \ref{sec:comparison} we perform comparative
studies of the galaxy populations in CGs, LGs and the field;
in section \ref{sec:sr} we compare some photometric scaling relations for galaxies in these
environments. We summarise and discuss our results in section \ref{sec:conc}. 
Throughout this paper we assume a flat cosmological model with parameters
$\Omega_0=0.3$, $\Omega_{\Lambda}=0.7$ and a Hubble's constant $H_0=100~h~\kms~{\rm Mpc}^{-1}$. 
All magnitudes have been corrected for Galactic extinction using the maps by \citet{sch98}
and are in the AB system. Absolute magnitudes and galaxy colours have been $K-$corrected using 
the method of \citet{Blanton:2003}~({\small KCORRECT} version 4.1).

\section{The samples}
\label{sec:sample}
\subsection{The sample of compact groups}
\label{subsec:grsample}

\begin{figure}[t]
\centering
\includegraphics[width=9cm]{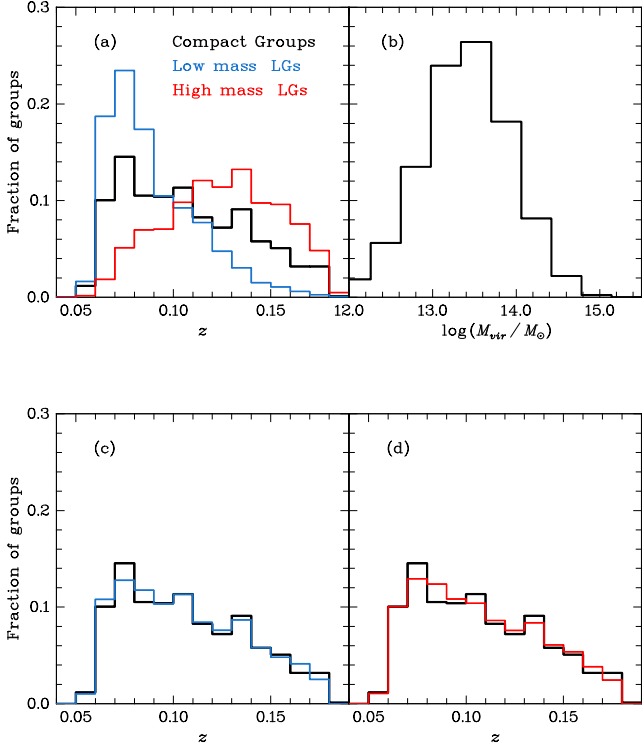}
\caption{ {\em Panel (a)} shows the spectroscopic redshift distribution of our samples of 
CGs ({\em black line}), and LGs with $N_{gal}\le 6$ in two 
ranges of virial mass: low-mass ({\em blue line}) and high-mass ({\em red line}).
{\em Panel (b)} shows the virial mass distribution of LGs groups with $N_{gal}\le 6$ 
and $0.06\le z\le0.18$. {\em Panels (c)} and {\em (d)} show the CGs redshift distribution 
({\em black line}), and the low-mass ({\em blue line}) and high-mass ({\em red line}) 
subsamples of LGs restricted to have redshift distribution similar to that of the 
CGs by using a Monte Carlo algorithm (see text for details).} 
\label{fig:dz}
\end{figure}

The sample of CGs used in this paper has been drawn from the catalogue of CGs identified by 
\citet{McConnachie:2009}. This catalogue was identified in the public release of the SDSS DR6 
(\citealt{dr6}). 
\citet{McConnachie:2009} used the original selection criteria of \citet{Hickson:1982}, 
which defined a CG as a group of galaxies with projected properties such that:
the number of galaxies within 3 magnitudes of the brightest galaxies is $N(\Delta m=3)\ge 4$;
the combined surface brightness of these galaxies is $\mu \le 26.0$ mag. arcsec$^{-2}$,
where the total flux of the galaxies is averaged over the smallest circle which contains
their geometric centres and has an angular diameter $\theta_G$; and $\theta_N\ge 3\theta_G$, 
where $\theta_N$ is the angular diameter of the largest concentric circle which contains
no additional galaxies in this magnitude range or brighter.

\citet{McConnachie:2009} identified 2,297 CGs, adding up to 9,713 galaxies down to a 
Petrosian (\citealt{petro76}) limiting magnitude of $r=18$ (Catalogue A), 
and 74,791 CGs (313,508 galaxies) down to a limiting magnitude of $r=21$ 
(Catalogue B). According to the authors, contamination due to 
gross photometric errors has been removed from the Catalogue A through 
the visual inspection of all galaxy members, and they estimated 
it is present in the Catalogue B at a 14\% level. The Catalogue A, which we use in this paper as a primary data source,
has spectroscopic information for 4,131 galaxies (43\% completeness). This catalogue includes groups
that have a maximum line-of-sight velocity difference smaller than $1000\, \kms$ only, to remove interlopers. 
The median redshift of the groups in this catalogue is $z_{\rm med}=0.09$. 
In this work we use a subsample of the Catalogue A of 
\citet{McConnachie:2009}, restricted to CGs in the redshift range $0.06\le z\le0.18$, 
which have spectroscopic redshift from at least one member galaxy, and also restricted 
our analyses to galaxy members with apparent magnitudes $14.5\le r\le 17.77$, i.e., the range in which
the Main Galaxy Sample (MGS; \citealt{Strauss:2002}) is complete. 
After meeting all these conditions, our group sample comprises 
846 CGs adding up to 2,270 galaxies, among which, 1,310 galaxies ($\sim 58\%$) have measured redshift. 
We show in the {\em panel (a)} of Fig. 
\ref{fig:dz} the redshift distribution of the CGs in our sample.
To every galaxy in the CGs with no redshift information we have assumed its redshift
to be the parent group's redshift.

\begin{figure}[t]
\centering
\includegraphics[width=7cm]{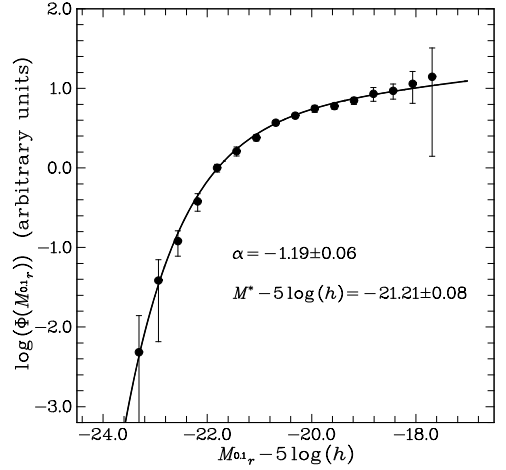}
\caption{The luminosity function of galaxies in Compact Groups. {\em Continuous line} is
the best fit Schechter function with shape parameters quoted inside the figure.
These parameters are used to compute the luminosities of CGs.}
 \label{fig:lf}
\end{figure}

\subsection{The sample of loose groups}
Loose groups of galaxies used in this paper are groups identified in redshift space,
and are not required to fulfil any compactness or isolation criterion. In particular,
we use groups drawn from the sample of \citet{ZM11} identified in the MGS of the Seventh 
Data Release (DR7, \citealt{dr7}).
Briefly, they used a friends-of-friends algorithm \citep{H&G:1982}
to link MGS galaxies into groups. This is followed by a second identification using
a higher density contrast on groups which have at least 10 members, in order
to split merged systems and clean up spurious member detection.
Given the known sampling problems for bright galaxies, the group identification was 
carried out over all MGS galaxies with $14.5 \le r \le 17.77$. 
Group virial masses were computed as $M = \sigma^2 R_{\rm vir} /G$, where $R_{\rm vir}$ 
is the virial radius of the system and $\sigma$ is the velocity dispersion of member
galaxies \citep{Limber:1960}. 
The velocity dispersion was estimated using the line-of-sight velocity dispersion $\sigma_v$ , 
$\sigma = \sqrt{3}\sigma_v$ .
The computation of $\sigma_v$ was carried out by the methods described by 
\citet{Beers:1990}, applying the biweight estimator for groups with
more than 15 member and the gapper estimator for poorer systems.
The sample of \citet{ZM11} (hereafter ZM11) comprises 15,961 groups which have more than 4 members,
adding up to 103,342 galaxies. Groups have a mean velocity dispersion of $193 \kms$, 
a mean virial mass of $2.1\times 10^{13} h^{-1} M_{\odot}$, and a mean virial radius of 
$0.9 h^{-1}\Mpc$. We refer the reader to ZM11 for details regarding
the identification procedure and parameters. 

\subsubsection{The high and low mass subsamples}
Since our work intends to perform a fair comparison between the galaxies inhabiting
CGs and LGs we do not use the groups in the ZM11 sample in a straightforward way.
It is well known that the properties of galaxies in
groups are correlated with group mass (e.g. \citealt{MM2:2006}), thus, in our analyses below
we compare galaxies in the CGs with galaxies in LGs in different mass rages.
We split groups in the ZM11 sample into 2 subsamples of low, 
$\log(M/M_{\odot}h^{-1})\le 13.2$, and high, $\log(M/M_{\odot}h^{-1})\ge 13.6$, mass.
These two subsamples have different redshift distributions, that also differ from that of 
the CGs, as can be seen in the {\em panel (a)} of Fig. \ref{fig:dz}.
Thus a direct comparison of the galaxies in these 2 subsamples and in the CGs 
will certainly be biased. We then use a Monte Carlo algorithm to randomly
select groups from these two subsamples in order to construct new subsamples
of low and high mass LGs that have redshift distributions similar to that of the
CGs. In {\em panels (c)} and {\em (d)} of Fig. \ref{fig:dz} we show the resulting redshift 
distributions and the redshift distribution of the CGs as a comparison. 
A Kolmogorov-Smirnov test (KS) among any of those distributions and that of the CGs 
gives significance levels for the null hypothesis that they are drawn from the same 
distribution above 95\%. Our final subsamples of low and high mass LGs
include 2,536 and 2,529 systems, adding up to 8,749 and 10,055 galaxies, respectively.

\subsubsection{The equal luminosity subsamples of compact and loose groups}

\begin{figure}[t]
\centering
\includegraphics[width=7cm]{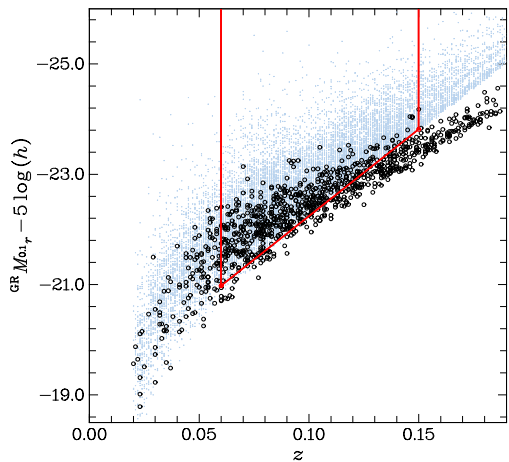}
\caption{The $^{0.1}r-$band group absolute magnitude ($^{GR}M_{^{0.1}r}$) as a function
of redshift. {\em Light blue dots} are the LGs, {\em open black circles} are the CGs. 
We show in {\em red lines} the region of the diagram within which we select subsamples
of LGs and CGs restricted to have similar redshift and absolute magnitude distributions.}
 \label{fig:select}
\end{figure}
\begin{figure}[t]
\centering
\includegraphics[width=9cm]{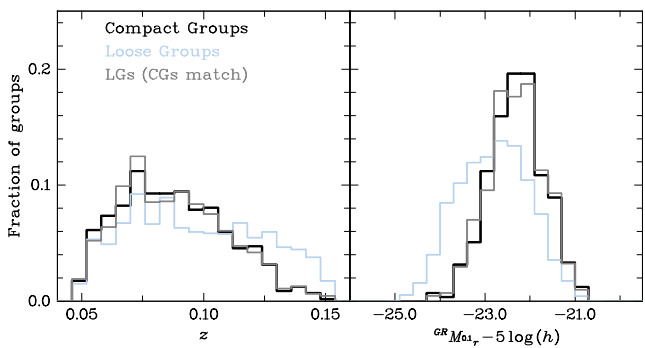}
\caption{Redshift ({\em left panel}) and total group absolute magnitude ({\em right panel}) 
distributions of groups in the region defined in Fig. \ref{fig:select}: 
CGs ({\em black}), LGs ({\em light blue})
and LGs selected by a Monte Carlo algorithm in order to have similar redshift and absolute
magnitude distributions as the CGs ({\em grey}).}
 \label{fig:LZ}
\end{figure}

Ideally, we would like to compare galaxies in LGs and CGs of similar masses, this can not be
done since CGs in the \citet{McConnachie:2009} catalogues do not have measured mass. Another way of comparing
CGs and LGs that have similar characteristics can be done by selecting them according to the total luminosity 
of their galaxy members. Thus, we also perform a comparison of galaxy properties in samples
of CGs and LGs with similar luminosity distributions. 
The luminosities of our LGs were computed by \citet{Martinez:2012}
by using the method of \citet{Moore:1993}, which accounts for the galaxy members not observed
due to the apparent magnitude limit by means of the luminosity function (LF) of galaxies
in groups. \citet{Martinez:2012} used the mass dependent LF 
of ZM11. 

To compute the total luminosities of the CGs using the method of \citet{Moore:1993}, we 
need first to compute the LF of the galaxies in CGs. 
We use two methods to compute the LF of galaxies in CGs: the non-parametric 
$C^{-}$ \citep{lb71,cho87} for the binned LF and the STY method \citep{sty} to compute the best fit 
\citet{Schechter:1976} function parameters. Since the Catalogue A of \citet{McConnachie:2009}
is complete down to an apparent magnitude $r=18$ we included all galaxies brighter
than this limit in the LF computation.
We show in Fig. \ref{fig:lf} the  resulting LF of galaxies in CGs
in the $^{0.1}r-$band. The best fitting Schechter function has shape parameters
$\alpha=-1.19\pm0.06$ and $M^{\ast}-5\log(h)=-21.21\pm0.08$ and clearly, is a good fit
to the $C^{-}$ points. 
It is interesting to compare this LF parameters with those found by ZM11 for the catalogue of LGs
used in this paper. On one hand, the $M^{\ast}$ is comparable with the value $-21.18\pm0.04$ that
ZM11 found for their highest mass bin, that is, groups with masses in the range 
$1.5-3.0 \times 10^{14}M_{\odot} h^{-1}$. 
On the other hand,
the faint end slope value is consistent with the value $-1.19\pm0.04$ corresponding to LGs of intermediate mass
$\sim 3.5\times 10^{13}M_{\odot} h^{-1}$ in ZM11.

\begin{figure}[t]
\centering
\includegraphics[width=7cm]{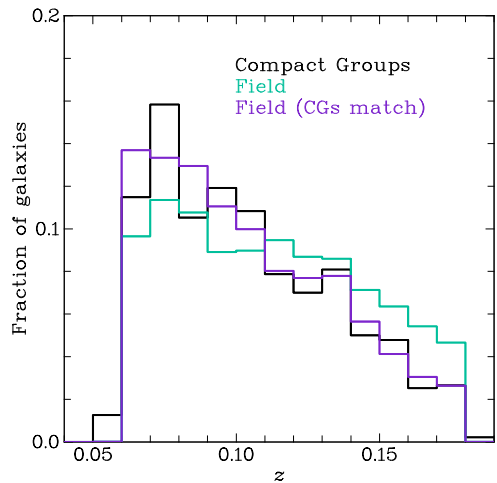}
\caption{Redshift distributions of galaxies: CGs sample ({\em black}), field galaxies ({\em green})
and field galaxies Monte Carlo selected to have a similar redshift distribution as CGs
({\em violet}).}
 \label{fig:dz2}
\end{figure}

We show in Fig. \ref{fig:select} the absolute magnitude of LGs and CGs as a function of redshift.
Clear differences are observed between LGs and CGs: at a fixed redshift the brightest objects are
LGs and the faintest are CGs. Among the LGs there are systems which can have more members than the 
typical CG and thus are brighter. CGs were identified over a parent catalogue which has a fainter
apparent magnitude limit ($r=18$) than the MGS ($r=17.77$), which is the parent catalogue 
for the LGs of ZM11. This explains why there are CGs much fainter than even the faintest LGs,
particularly at $z>0.15$.
\begin{figure*}[t]
\centering
\includegraphics[width=18cm]{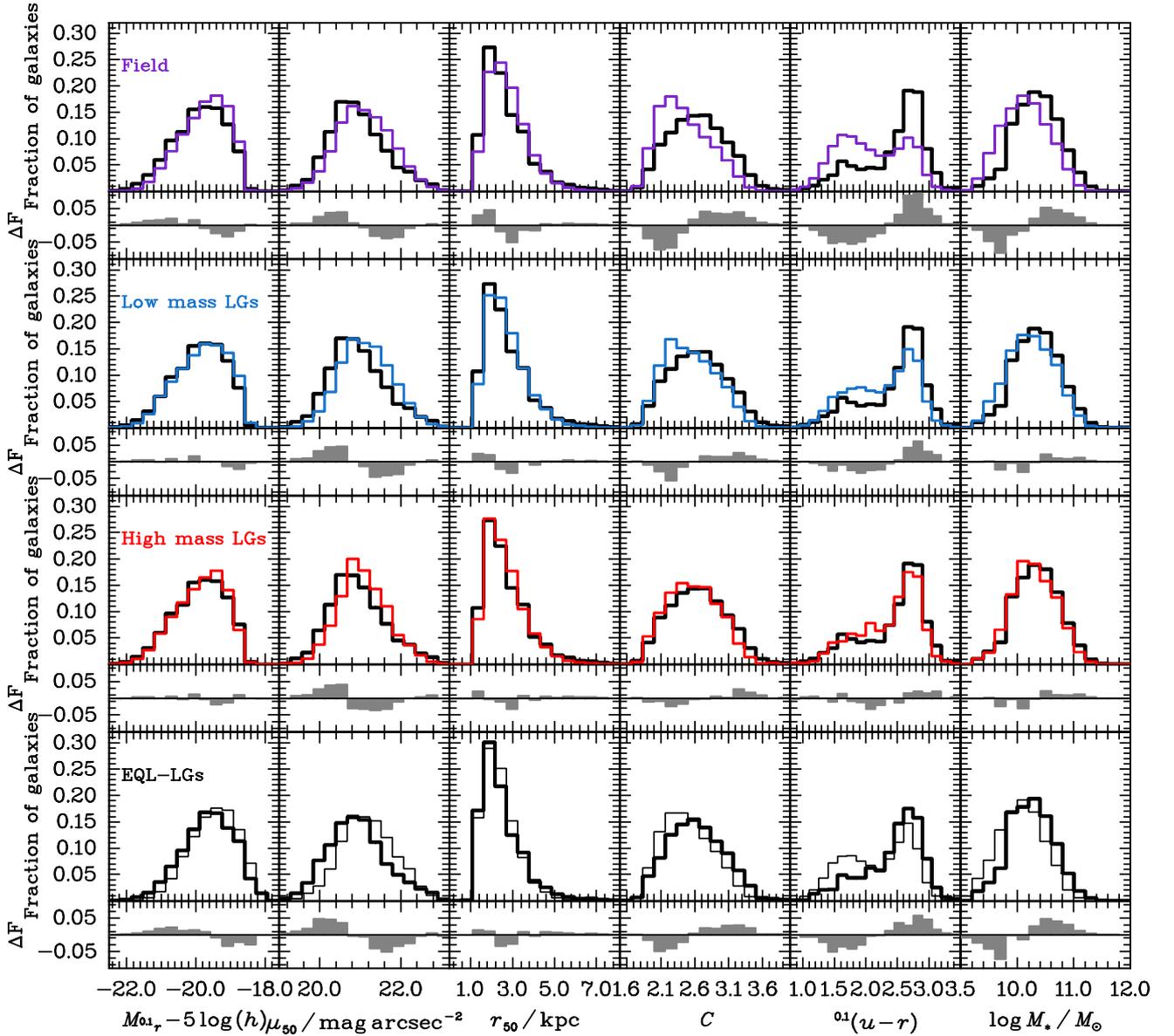}
\caption{Distributions of galaxy properties in our samples: CGs sample ({\em thick black line}), 
field ({\em violet}), low-mass LGs ({\em blue}), high-mass LGs ({\em red}) and EQL-LG 
({\em thin black line}).
All distributions have been normalised to have the same area.
Below each panel we show as {\em shaded histograms} the residuals between the distributions.}
\label{fig:histo}
\end{figure*}

To perform a fair comparison between LGs and CGs having similar absolute magnitude distributions
we firstly define a region in the redshift-absolute magnitude plane in which we can find 
both types of groups. 
As discussed above, for all redshifts $z>0.05$, the CG sample includes systems which are fainter
than all the LGs due to differences in the apparent magnitude limit of the parent catalogues. 
As can be seen in Figure \ref{fig:select} beyond $z>0.15$ the samples
of LGs and CGs do not overlap at all, this imposes the upper redshift cut-off. For the lower
redshift cut-off we use the same value, $z=0.06$, as in the other samples defined before.
Within this redshift range, we avoid the region in which only CGs are found, that is, we
impose a redshift dependent faint absolute magnitude cut-off which, for simplicity, we have chosen
to be linear. We indicate this region with {\em red lines} in Fig. \ref{fig:select}.
Now, within this region, LGs and CGs still have different redshift and absolute magnitude distributions
which we show in Fig. \ref{fig:LZ}  as {\em black} and {\em light blue} histograms respectively.
We then select by means of a Monte Carlo algorithm, a subsample of LGs that match the redshift
and absolute magnitude distributions of the CGs. This is shown as {\em grey} histograms in
Fig. \ref{fig:LZ}. When we compare LGs and CGs of equal luminosity in the
analyses below, we refer to these two subsamples of groups. Our final subsamples of EQL-CGs and
EQL-LGs include 571 and 2,345 systems, adding up to 1,729 and 10,554 galaxies, respectively.
We explicitly make this distinction between the CG and the EQL-CG samples since the latter has a redshift
dependent absolute magnitude constraint that is absent in the former.

\subsection{The sample of field galaxies}
\label{subsec:galsample}

In this work we also compare the properties of galaxies in CGs with the properties of field galaxies.
We consider as field galaxies to all DR7 MGS galaxies that were not identified as belonging to 
LGs by ZM11 groups or to CGs by \citet{McConnachie:2009}, and have apparent magnitudes
$14.5\le r\le 17.77$.
For an adequate comparison with our samples of galaxies in groups, we used the same Monte Carlo
algorithm of the previous subsection to contract a sample of field galaxies that has a similar
redshift distribution as that of galaxies in our CG sample. This field sample includes 250,725
galaxies. We show in Fig. \ref{fig:dz2} the redshift distributions of galaxies in CGs, of all 
field galaxies and of field galaxies Monte Carlo selected. 
A KS test between the galaxies in CGs and in our Monte Carlo
selected field samples gives significance levels for the null hypothesis 
that they are drawn from the same distribution above 95\%. 
From now on, when we refer to field galaxies we mean galaxies in this Monte Carlo selected
sample.

\section{Comparing galaxies in CGs, LGs and in the field}
\label{sec:comparison}

In this section we compare parameters of galaxies in CGs, LGs and in the field.
The parameters we have focused our study on are: 
\begin{itemize}
\item Petrosian absolute magnitude in the $^{0.1}r-$band;
\item The radius that encloses 50\% of the Petrosian flux $r_{50}$;
\item The $r$-band surface brightness, $\mu_{50}$, computed inside $r_{50}$;
\item The concentration index, defined as the ratio of the radii enclosing
90 and 50 percent of the Petrosian flux, $C = r_{90}/r_{50}$;
\item The $^{0.1}(u-r)$ colour. We use model instead of Petrosian magnitudes to compute colours
since aperture photometry may include non-negligible
Poisson and background subtraction uncertainties in the $u$ band;
\item The stellar mass, $M_{\ast}$ based on luminosity and colour, 
computed following \citet{Taylor:2011}.
\end{itemize}

In the analyses below, we classify galaxies into early and late types according 
to their concentration index. Typically, early-type galaxies have $C > 2.5$, while for 
late-types $C < 2.5$ (\citealt{Strateva:2001}). The effects of seeing in the measurement of 
$r_{50}$ and $r_{90}$ have to be considered for galaxies with relatively small
angular size. The average seeing in the SDSS is below a conservative
value of $1.5\arcsec$ \citep{Shen:2003}. Since the values of $r_{50}$, $\mu_{50}$ and $C$ can be
unreliable for galaxies with $r_{50}$ below this value, we have excluded them
from our analyses. The numbers of galaxies in each sample quoted in the previous section
already contemplate this size cut-off.

Since many galaxy properties correlate with absolute magnitude, thus, to perform a fair comparison
we weight each galaxy in our computations by $1/V_{max}$ \citep{Schmidt:68} in order to 
compensate for the fact that we are dealing with galaxy samples that are drawn from flux-limited
catalogues. Fig. \ref{fig:histo} compares the normalised distributions of galaxy parameters of galaxies
in LGs and in the field with those of CGs galaxies. 
Below each panel of Fig. \ref{fig:histo} we show the residuals between each pair of distributions, i.e.,
for each property $X$, the difference $\Delta F(X)=f_{CG}(X)-f(X)$, where $f_{CG}(X)$
and $f(X)$ are the fractions of galaxies in the bin centred on $X$ in the CGs and in
the other sample, respectively. 

Regarding the luminosity, and as can be seen from Fig. \ref{fig:histo}, galaxies in CGs tend to be 
slightly more luminous than their field counterparts, in the sense of an excess of
$M_{^{0.1}r}-5\log(h)\lesssim -20$ galaxies, this is in agreement with previous findings such as
\citet{Deng:2008}. We find no clear difference with either low, and high mass LGs.
A similar result is found by \citet{Deng:2007a} when comparing galaxies in CGs and LGs
identified by different algorithms. Important differences can be seen between CGs and the
sample of LGs restricted to have similar total luminosity distribution (hereafter EQL-LG):
galaxies in CGs are systematically brighter. 

Compared to all LG samples and the field, CGs have a larger fraction of galaxies with 
$\mu_{50}\lesssim  20.4 \,{\rm mag\, arcsec^{-2}}$ and a deficit of lower surface brightness galaxies. 
\begin{figure}[t]
\centering
\includegraphics[width=9cm]{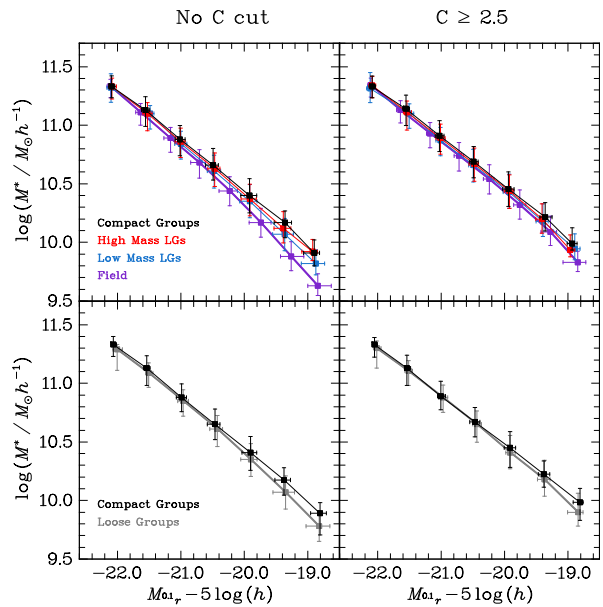}
\caption{The stellar mass as a function absolute magnitude for galaxies in our samples. 
Points represent the median
in each bin, error-bars are the 25 and 75\% percentile within each bin.
}
\label{masslum}
\end{figure}

When comparing galaxy sizes, we find differences between CGs and the other environments
for galaxies with $r_{50}\lesssim 3 \,{\rm kpc}$. CGs have an excess of galaxies with
$r_{50}\lesssim 2 \,{\rm kpc}$ and a deficit of $2\,{\rm kpc}\lesssim r_{50}\lesssim 3 \,{\rm kpc}$
galaxies. \citet{Deng:2008} do not find significant differences in the sizes of galaxies in CGs and 
a sample of field galaxies, which they argue it is due to their narrow luminosity range.

Galaxies in CGs are systematically more concentrated than their counterparts
in the field or in LGs. This difference mirrors the fact that galaxies in CGs have, on average, smaller
sizes and not very different luminosities than galaxies in the other samples.
Thus, CGs have a higher fraction of early-type galaxies. In agreement with our results,
\citet{Deng:2008} find that CGs have a larger fraction of high concentrated early-type galaxies
when compared to field galaxies.

In agreement with the excess of early type galaxies, galaxies in CGs show a higher fraction 
of red galaxies, compared to the field and the LGs. 
Our results agree with the comparison of CGs and field galaxies
by \citet{Lee:2004} and \citet{Deng:2008}. \citet{Brasseur:2009} find similar 
results performing a similar comparison using mock catalogues based upon the 
Millennium Run simulation \citep{Springel:2005}.

Galaxies in CGs tend to have larger stellar masses than their field and EQL-LGs counterparts.
We further explore this in Fig. \ref{masslum}, where we show the median stellar mass as a function
of absolute magnitude for galaxies in all our samples. Differences arise at the lower luminosities
we explore, galaxies in groups differ from field galaxies, being more massive at fixed luminosity.
At the same time something similar is observed when comparing the equal luminosity subsamples:
there is a hint for galaxies in CGs to be more massive at lower luminosities. All these differences
are almost erased when we consider early types only. 

As a general conclusion from this section, galaxies in CGs differ from galaxies
in other environments. Differences are larger when compared to field galaxies and smaller
when compared to galaxies in high mass LGs. 
\begin{figure*}[t]
\centering
\includegraphics[width=12cm, angle=270]{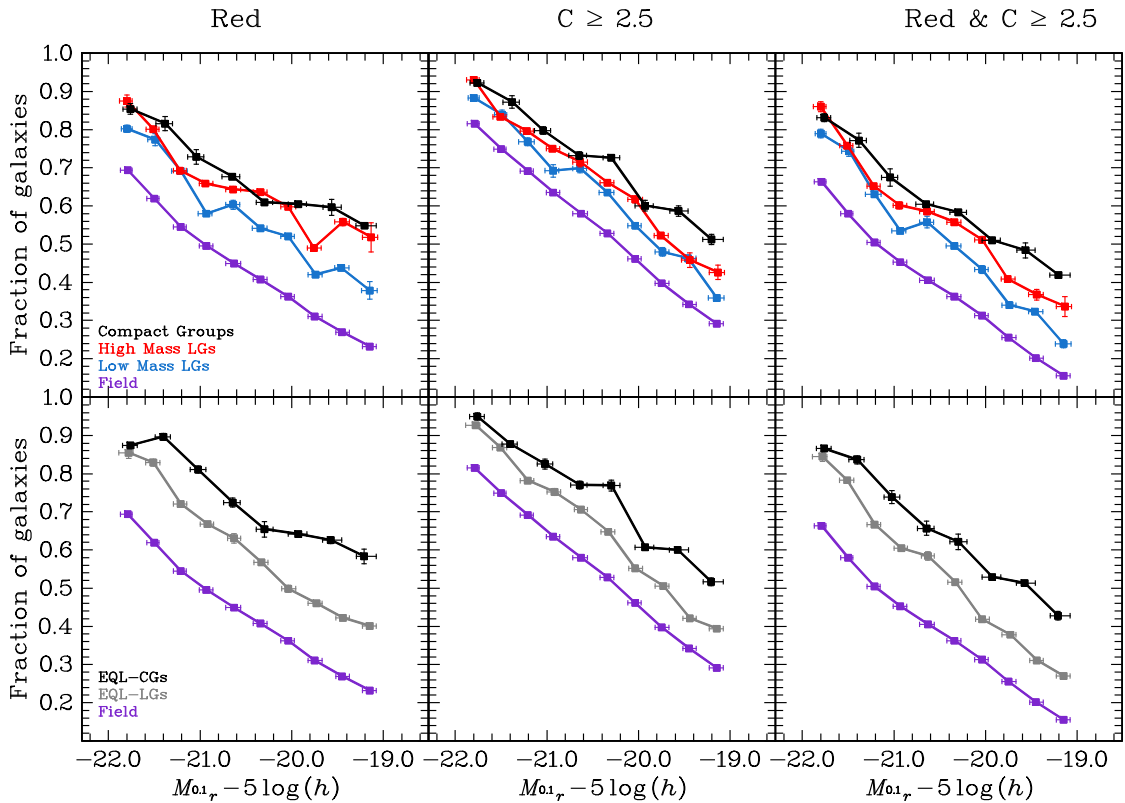}
\caption{{\em Left panels:} the fraction of red galaxies according to their $^{0.1}(u-r)$ colour; 
{\em centre panels:} the fraction of early-type galaxies according to their concentration parameter;
 {\em right panels:} the fraction of red early-type galaxies. All fractions are shown
as a function of absolute magnitude.
The {\em upper panels} compare CGs, LGs of low and high mass and field galaxies, 
while {\em lower panels} compare CGs and LGs with similar redshift and total absolute 
magnitude. 
Vertical error-bars are obtained by using the bootstrap resampling technique,
horizontal error-bars are the 25 \& 75\% quartiles of the absolute magnitude distribution within each bin.
}
\label{fig:frac1}
\end{figure*}
\subsection{The fraction of red and early type galaxies in groups}

As a complementary study, we also study the fraction of galaxies that are in the red sequence or 
are classified as early type as a function of galaxy absolute magnitude in the environments we probe.
To quantify the fraction of red galaxies, we follow ZM11 and classify galaxies 
as red/blue accordingly to whether their $^{0.1}(u-r)$ colour is larger/smaller 
than the luminosity dependent threshold 
$T(x)=-0.02x^2-0.15x+2.46$, where $x=M_{^{0.1}r}-5\log(h)+20$. 
To classify galaxies into early and late, we use the concentration parameter and consider as early types
those galaxies which have $C>2.5$.

Fig. \ref{fig:frac1} shows the fraction of red galaxies (\textit{left panels}),
early type galaxies (\textit{centre panels}) and red early-types galaxies (\textit{right panels})
as a function of galaxy absolute magnitude. The comparison among CGs, LGs samples and the field shows that
CGs have a larger fraction of red early-type galaxies over the whole absolute magnitude range. 
For brighter luminosities, CGs and high mass LGs have similar fraction of red galaxies, the
largest difference is observed when red and early-types galaxies are considered.
Important differences can be seen between CGs and LGs of similar luminosities ({\em lower panels}).

\section{Photometric relations}
\label{sec:sr}
There are well known scaling relations involving photometric, structural and dynamical parameters of 
galaxies. Among the scaling relations that involve photometric parameters are the 
colour-magnitude \citep{SV:1978b,SV:1978a}, also known as the red sequence (RS) for early-type galaxies,
and the luminosity-size relation.
These empirical relations are closely related to the physical processes involved in the galaxy 
formation scenario and, therefore, are a fundamental tool to understand the formation and 
evolution of galaxies. In this section, we compare these relations for our samples of
galaxies in CGs, LGs and in the field.
As in the previous section we weight each galaxy by $1/V_{max}$ according to its absolute magnitude
and the redshift and apparent magnitude cut-offs.

\begin{figure}[t]
\centering
\subfigure{\includegraphics[width=9cm]{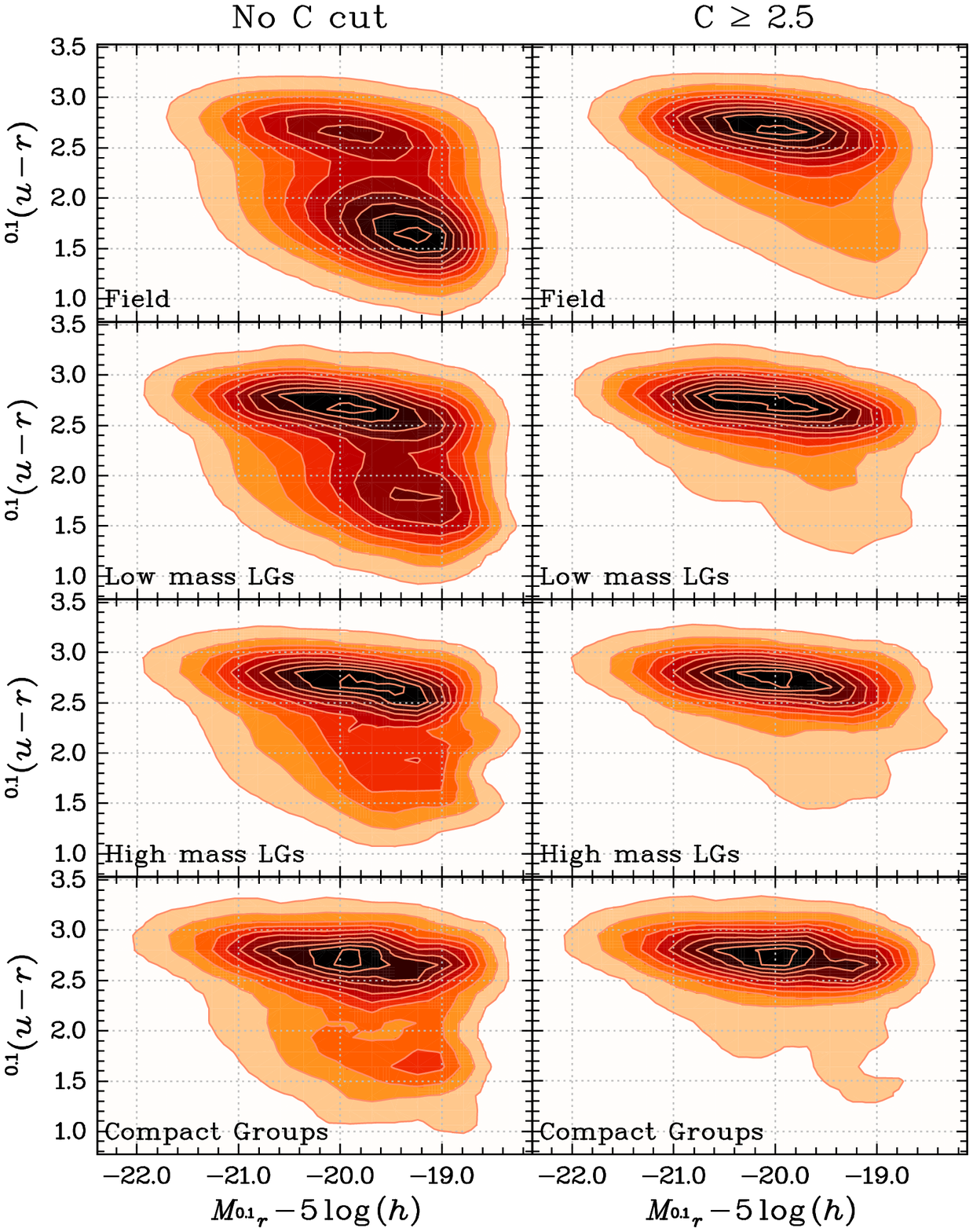}}
\subfigure{\includegraphics[width=9cm]{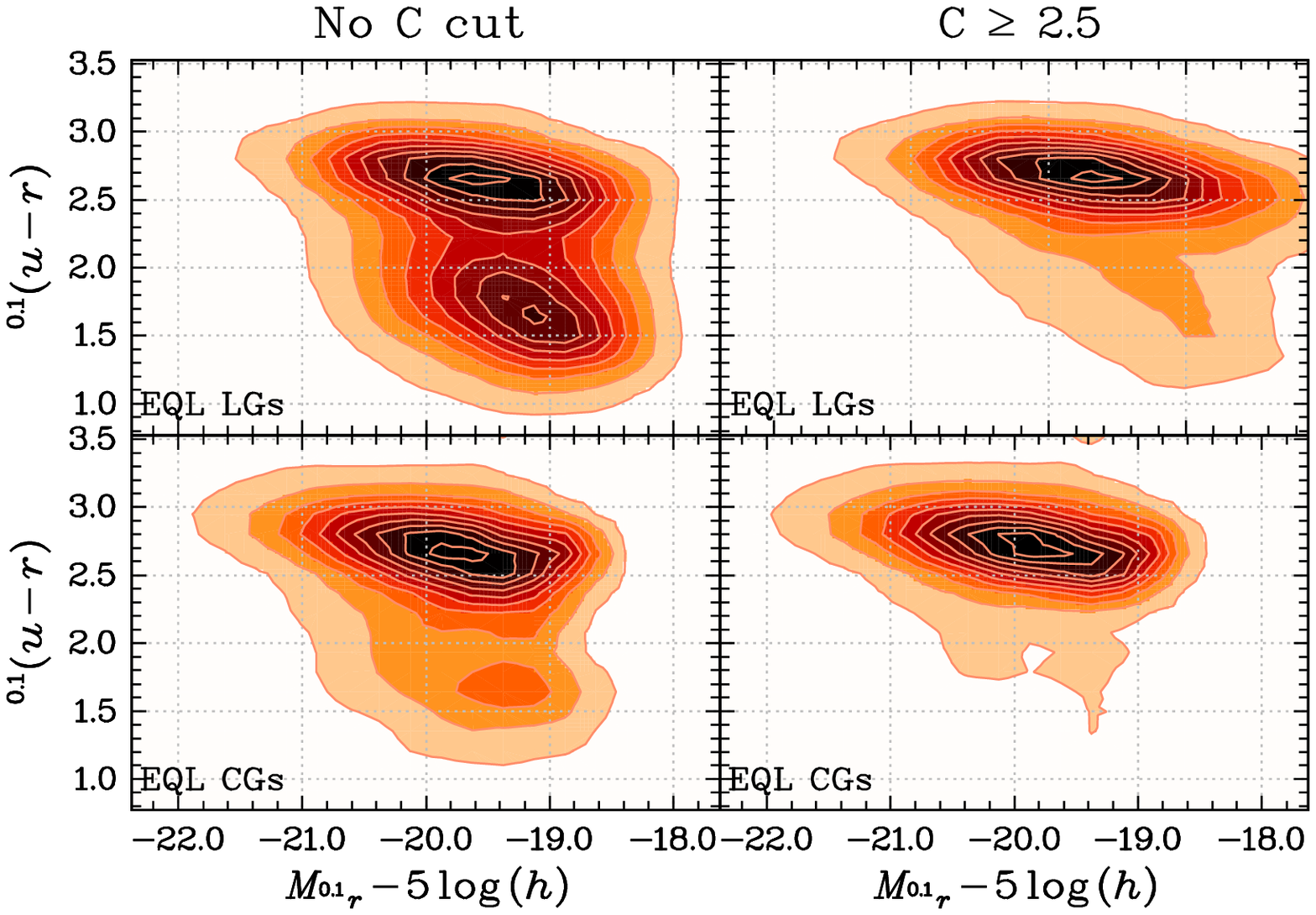}}
\caption{
Colour-magnitude diagram: $^{0.1}(u-r)$ as a function of $M_{^{0.1}r}$ for galaxies in the field, 
in low and high mass LGs, and in CGs. Equal luminosity samples of CGs and LGs are
shown separately.
{\em Left panels} include all galaxies in the samples, 
while {\em right panels} include only early-type galaxies according to their concentration parameter.
{\em Darker colours} represent higher values.
All distributions have been normalised to enclose the same volume.}
 \label{fig:rs_con}
\end{figure}
\begin{figure}[t]
\centering
\includegraphics[width=9cm]{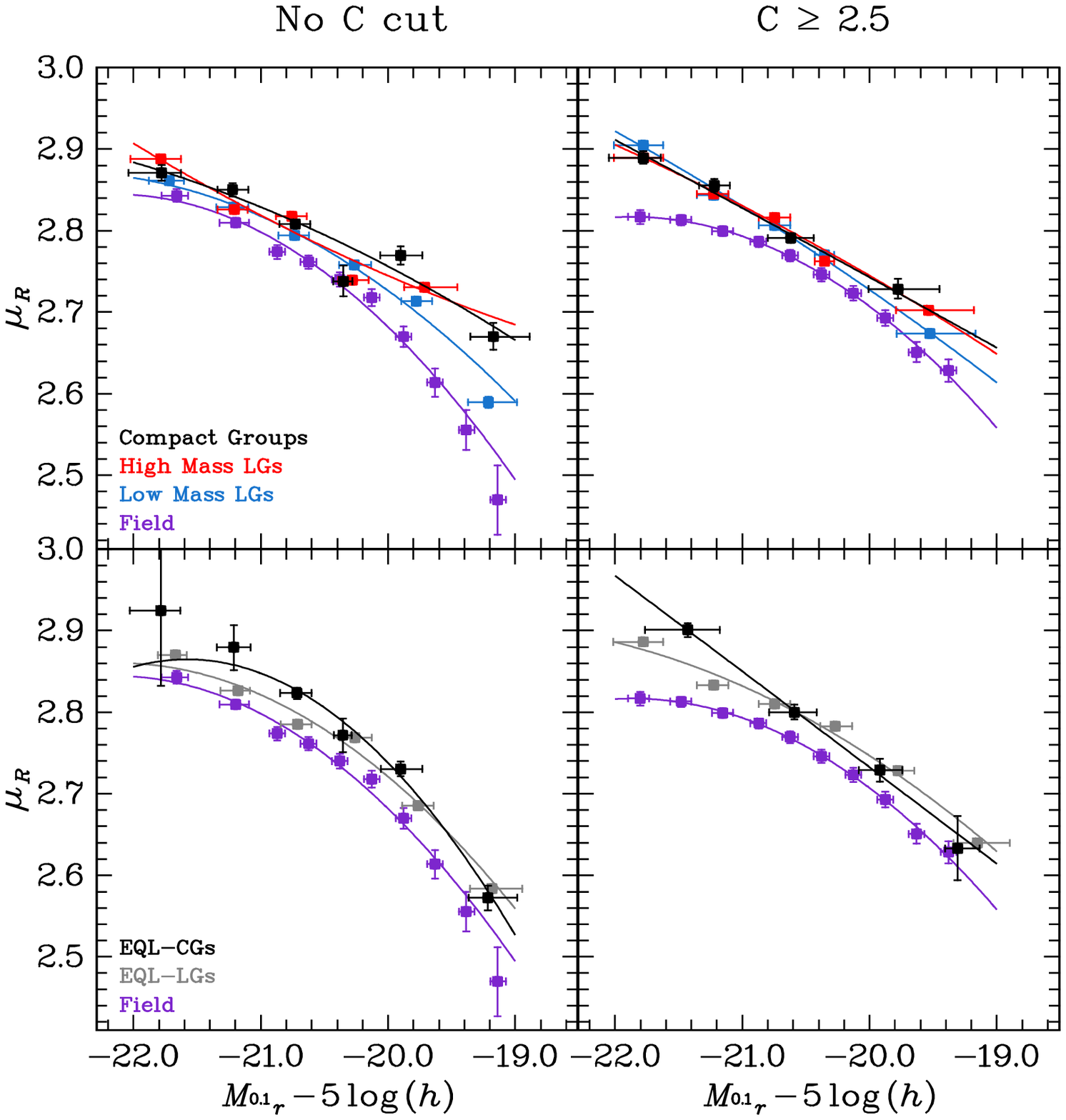}
\caption{$\mu_R$ of the sequences of red galaxies from Fig. \ref{fig:rs_con}  
as a function of absolute magnitude. The abscissas are
the median and the horizontal error bars are the $25$ and $75$ \% quartiles
of the absolute magnitude distribution within each luminosity bin. Vertical
error bars are the $1\sigma$ error estimates from the fitting procedure.
Continuous lines are the best-fitting quadratic models.
}
\label{fig:rs_sec}
\end{figure}

\subsection{Colour-magnitude diagram: red sequence}
\label{ssec:rs}

Fig. \ref{fig:rs_con} shows the colour-magnitude diagram as a function of environment.
As expected, it is clear from the {\em left panels} of Fig. \ref{fig:rs_con}, that, as we move from field to LGs 
of increasing mass, the blue population declines in numbers, while red galaxies become the dominant population.
For CGs the red population is even more dominant. The {\em right panels} of Fig. \ref{fig:rs_con} 
consider only early-type galaxies, this does not completely
remove a blue population which also becomes less prominent as we move from field to CGs.  
\citet{Lee:2004} and \citet{Brasseur:2009} found that galaxies in compact associations are 
confined nearly exclusively to the red sequence, with few galaxies occupying the blue cloud, 
in agreement with our results. 

It is well known that, for a fixed absolute magnitude, $M$, the colour distribution of the galaxies 
is well described by the sum of two Gaussian functions representing the blue cloud and the 
red sequence (e.g. \citealt{Baldry:2004,Balogh:2004,Martinez:2006}):
\begin{equation}
f(col|M)=A_B\exp\left(-\frac{(col-\mu_B)^2}{2\sigma_B^2}\right)
+A_R\exp\left(-\frac{(col-\mu_R)^2}{2\sigma_R^2}\right).
\label{2G}
\end{equation}

In the equation above, $A_B$ and $A_R$ are the amplitudes, $\mu_B$ and $\mu_R$ the centres, 
$\sigma_B$ and $\sigma_R$ the width of the Gaussian functions describing the blue ($B$) 
and red ($R$) populations.
We study the environmental dependence of the red sequence following the same procedure 
as in \citet{Martinez:2010}: for different absolute magnitude bins, we fit the two Gaussian model 
(Eq. \ref{2G}) to the $^{0.1}(u-r)$ colour distribution of field, LG and CG galaxies, using a
standard Levenberg-Marquardt method. Thus, for all our galaxy samples we have the 6 parameters of
Eq. \ref{2G} as a function of $^{0.1}r-$band absolute magnitude. 
In Fig. \ref{fig:rs_sec} we show the centre of the Gaussian function ($\mu_R$)  that
fits best the red sequence in the colour-magnitude diagrams of Fig. \ref{fig:rs_con}.
The abscissas are the medians of the corresponding distributions of the absolute
magnitudes in each bin and the horizontal error bars are the $25$ and $75$ \% quartiles.

In the absolute magnitude range we probe in this work, we find that data points in 
Fig. \ref{fig:rs_sec} are well described by a quadratic polynomial \citep{Martinez:2010}. 
The continuous lines in figure \ref{fig:rs_sec} show the best quadratic fits to
$\mu_R$ as a function of the absolute magnitude. This figure shows ({\em upper left panel})
that the red sequence of field galaxies is always bluer than its counterparts in groups.
Among groups, the mean colour of the red sequence 
is systematically redder for the high mass subsample (as in \citealt{Martinez:2010}). 
The $\mu_R$ of CG galaxies is, consistent with that of galaxies in high mass LGs over the
whole range of absolute magnitudes we probe. When considering the samples of CGs and LGs
of similar luminosities ({\em lower left panel}), 
$\mu_R$ of CGs is systematically redder for $M_{^{0.1}r}-5\log(h)<-20.5$.  
When we consider red early-type galaxies alone ({\em right panels} of Fig. \ref{fig:rs_sec}), the differences
between field, LGs and CGs disappear. Nevertheless, differences between 
the red sequences of CGs, LGs and the field are still present.

\subsection{Luminosity-size relation}
\label{ssec:sr}

\begin{figure}[t]
\centering
\subfigure{\includegraphics[width=9cm]{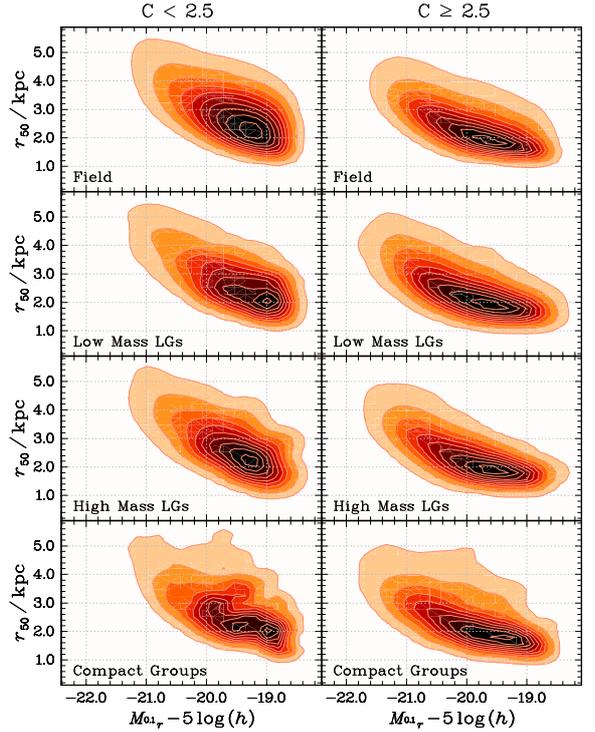}}
\subfigure{\includegraphics[width=9cm]{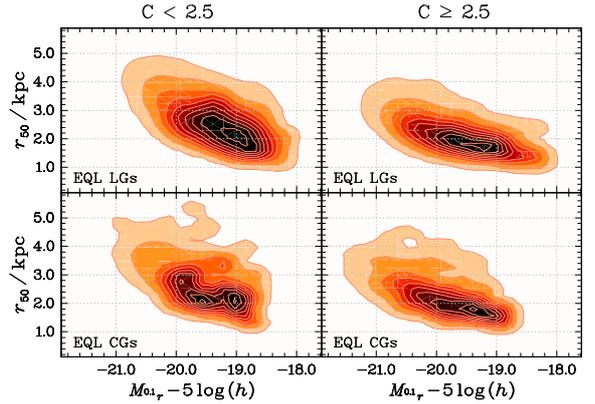}}
\caption{Petrosian half-light radius, $r_{50}$, as function of the absolute magnitude 
for field, low and high mass LG and CG galaxies. Equal luminosity samples of CGs and LGs are
shown separately.
{\em Left panels} show late-type galaxies, while {\em right panels} consider early-type galaxies according
to their concentration parameter.} 
 \label{fig:lz_con}
\end{figure}

Fig. \ref{fig:lz_con} shows the Petrosian half-light radius as a function of absolute magnitude 
of late ({\em lefts panels}) and early-type galaxies ({\em right panels}). 
As it is expected, brighter galaxies
are larger, as has been found in several environments and regardless morphological types 
(e.g. \citealt{Coenda:2005,Bernardi:2007,vonderlinden:2007,Coenda:2009,Nair:2010}). 

\begin{figure}[t]
\centering
\includegraphics[width=9cm]{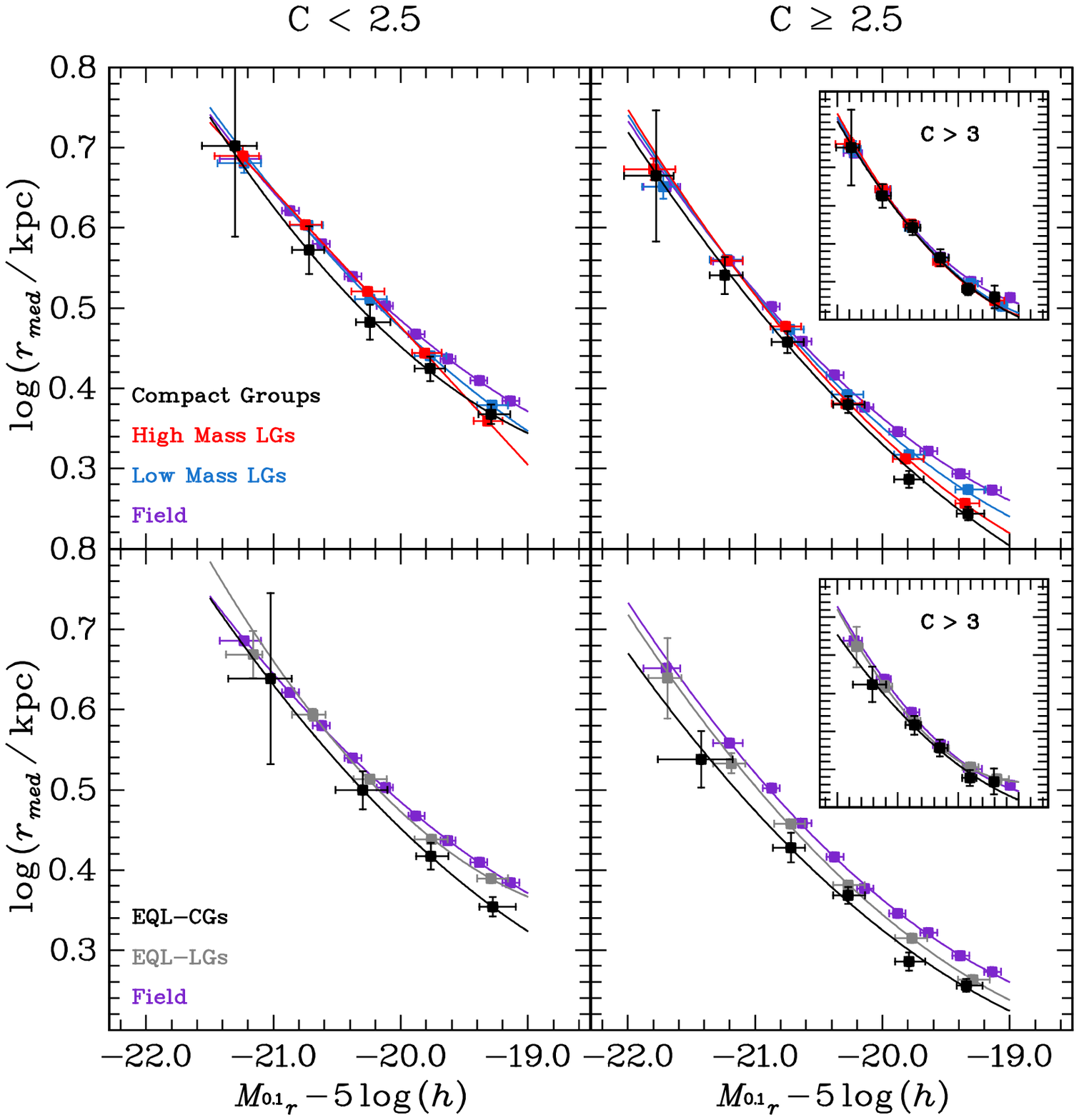}
\caption{The $\log({r}_{\rm med})$ of the size distribution as a function of absolute magnitude 
for late-type galaxies 
({\em left panels}) and early-types galaxies ({\em right panels}). The abscissas are
the median and the horizontal error bars are the $25$ and $75$ \% quartiles
of the absolute magnitude distribution within each luminosity bin. Vertical
error bars are the $1\sigma$ error estimates from the fitting procedure.
Continuous lines are the best-fitting quadratic models.
{\em Inset panels} show the results of using $C>3$ to define early-type galaxies. 
}
 \label{fig:lz_seq}
\end{figure}

To analyse the luminosity-size relation, for each panel in Fig. \ref{fig:lz_con} we derive the size 
distribution within several absolute magnitude bins. In all cases, for a fixed absolute magnitude, $M$,
the size ($r_{50}$) distribution can be well described by a log-normal distribution \citep{Shen:2003}, 
which is characterised by a median ($\mu=\ln(r_{\rm med})$) and a dispersion ($\sigma$) 
\begin{equation}
f(r_{50}|M)=\frac{1}{r_{50}\sigma \sqrt{2\pi}}\exp\left(-\frac{\ln(r_{50})-\mu}{2\sigma^2} \right).
\label{LGN}
\end{equation}
We fit this model to the distributions shown in Fig. \ref{fig:lz_con} using a standard Levenberg-Marquardt
procedure. Fig. \ref{fig:lz_seq} shows the median $\log({r}_{\rm med})$ 
as a function of absolute magnitude of 
the distributions shown in Fig. \ref{fig:lz_con}. For all samples of galaxies analysed here, 
a quadratic polynomial function is a good description of the median $\log({r}_{\rm med})$  as a function
of absolute magnitude. We also show these fits in Fig. \ref{fig:lz_seq}.  
The curvature in the luminosity-size relation has been previously reported by \cite{Bernardi:2007} 
and \citet{Coenda:2009}.

Taking into account error-bars, the differences among the different sequences in Fig. 
\ref{fig:lz_seq} are not significant for most of the bins. The only clear difference is
seen between early-type galaxies populating the EQL CG sample and the field.
Nevertheless, a systematic behaviour can be seen in all panels of Fig. \ref{fig:lz_seq}: over 
the whole range on luminosities, galaxies in CGs tend to be the smallest, 
while field galaxies are the largest ones. 
This effect is observed for both, early and late-type galaxies. \citet{Weinmann:2009} analysing 
central and satellite galaxies found a similar dependence of the size-luminosity relation with the 
environment, nevertheless, the effect is only observed for late-type galaxies. These authors adopted 
a more restrictive $C>3$ value to select early-type galaxies. This threshold preferentially selects 
elliptical galaxies. In order to compare our results with those obtained by \citet{Weinmann:2009} we have 
considered a sub-sample of early-type galaxies assuming a $C>3$. 
The corresponding results are shown in 
the inset-panels of Figure \ref{fig:lz_seq}, where it can be observed that the size-luminosity relation for $C>3$ 
galaxies is the same for all the environments considered. This is also in agreement with 
\citet{Nair:2010} who, by using a sample of visually classified bright sample of galaxies,
found no dependence with environment of the size-luminosity relation for elliptical galaxies.

\section{Conclusions and Discussion}
\label{sec:conc}

To investigate the dependence of the galaxy properties on environment, we performed a comparative
study of the properties of galaxies in CGs, LGs and in the field in the redshift range $0.06<z<0.18$.
CGs used in this paper were drawn from the Catalogue A of \citet{McConnachie:2009},
while LGs were selected from the sample of ZM11.
In all cases, galaxy properties used in our work were taken
from the MGS sample of the SDSS DR7.

We selected three samples of LG  taken from the ZM11
catalogue: low ($\log(M/M_{\odot}h^{-1})\le 13.2$) and high ($\log(M/M_{\odot}h^{-1})\ge 13.6$)
mass. The third sample was selected in order to have similar total luminosity
distribution than that of CGs. Since the original samples of CGs and LGs have
different redshift distributions, we constructed the LGs group samples by using a
Monte Carlo algorithm which randomly selects groups in order to
reproduce the redshift distribution of CGs.
Similarly, our sample of field galaxies was drawn
to reproduce the redshift distribution of CG members. The final
samples have 846, 2,536, 2,529 of compact, low-mass, high-mass
and equal luminosity loose groups respectively.
The corresponding number of member galaxies are: 2,270, 8,749 and 10,055.
The equal luminosity subsamples of compact and loose groups include 571 and 2,345 objects,
adding up to 1,729 and 10,554 galaxies respectively.
The field sample comprises 250,725 galaxies. 
This statistically significant set of data has
been used to compare basic properties of galaxies as well as some
photometric scaling relations
in different environments.

Our main findings are:
\begin{itemize}
\item The properties of galaxies in LGs or in the field do not match
those of galaxies in CGs.
\item Compact groups are the environment that shows the highest
fraction of early-type and red galaxies (our comparison between CGs and field 
agrees with \citealt{Lee:2004,Deng:2008,Brasseur:2009}).
This effect is observed for the whole range of absolute magnitude and
stellar mass. 
\item Galaxies in CGs are, on average, smaller, more compact and have
higher surface brightness and stellar mass than in LGs or in
the field. Differences are larger when compared to field galaxies and smaller when compared
to galaxies in high mass LGs. This disagrees with previous findings by \citet{Deng:2008},
but it should be kept in mind they explore a narrower range in luminosity.
\item The luminosity function of galaxies in CGs has a characteristic magnitude comparable
to that of the most massive LGs, while its faint end slope is similar to that of LGs of intermediate mass
(LG luminosity functions measurements by ZM11).
These parameters might be indicating that the compact group environment is effective in producing
bright galaxies and, at the same time, is a more hostile environment for fainter galaxies compared to
LGs. Nevertheless, solid conclusions on this will be obtained when mass measurements of CGs 
allow a more detailed study of their LF and its dependence on mass.
\item The mean colour of CG galaxies is consistent with that of galaxies
in high mass LGs over the whole range of absolute magnitudes we probe.
\item For a fixed luminosity and over the whole range of absolute
magnitudes, both, late and early-type galaxies in CGs are smaller than in EQL groups and in the field.
A similar trend is observed when compared to galaxies in low mass groups, although it is not statistically 
significant. If early-type galaxies are selected using $C>3$, the corresponding size-luminosity 
relations are environment independent, in agreement with \citet{Weinmann:2009}, \citet{Nair:2010}
and \citet{Maltby:2010}.
\end{itemize}

It should be taken into account that we have excluded from our analyses galaxies
with $r_{50}$ below the average seeing in SDSS images. While this
avoids introducing
systematics due to the seeing, it also excludes increasingly larger galaxies as
we go from the smallest to the largest redshift considered in this work. Thus
the actual differences between galaxies in the different environments probed
here might be more significant.

Our results do not significantly change if we: {\em (i)} consider only
galaxies with spectroscopic
redshift in CGs; {\em (ii)} restrict our analyses to CGs with higher values
of surface brightness. We refer the reader to Appendix \ref{sec:appx} below.

One of our most important results is the excess of galaxies in CGs that are
more compact, redder and have higher surface brightness with respect
to their LG or field counterparts.
These galaxies could be the descendants of galaxies that inhabited LGs
before going through a phase of CG. In the CG environment, galaxies
have undergone mergers and tidal effects caused by the high densities
and low velocity dispersions that characterise CGs. This is in agreement with the large
number of CGs that show clear evidences of disturbed morphology (e.g.
\citealt{MOH:1994}). The high fraction of red galaxies suggests that this is an efficient process
ending in objects with earlier morphological types. This high fraction of red galaxies
in CGs also evidence an
advanced stage of the morphological transformation processes, consistent with
predictions using numerical simulations, see for instance
\citet{Brasseur:2009}, who conclude that galaxies in CGs should be mainly red and dead ellipticals.
Our results are also consistent with studies of SFR in compact groups, such as 
\citet{Walker:2010} and \citet{Tzanavaris:2010}.
The differences between the luminosity function of galaxies in CGs and
LGs also support this scenario where low luminosity galaxies merge efficiently
producing both a lower number
of faint galaxies and higher number of bright early-type galaxies as observed.

Recent results (e.g. \citealt{Cortese:2006,Wilman:2009}, \citet{McGee:2009}
suggest that groups of galaxies play a fundamental role in the pre-processing of galaxies before they
become part of more massive systems like clusters of galaxies. Our results indicate that
galaxies that inhabit a high density environment, like that of the CGs, have
undergone a major transformation than those objects that just went through a phase of LG.
Within this scenario, the properties of galaxies in high mass systems
like clusters, should show
cosmic variations in the galaxy properties depending on the fraction
of members that went through
a phase of CG.

\begin{appendix}
\section{The Compact Groups Sample: Contamination}
\label{sec:appx}

\begin{figure*}[t]
\centering
\includegraphics[width=18cm]{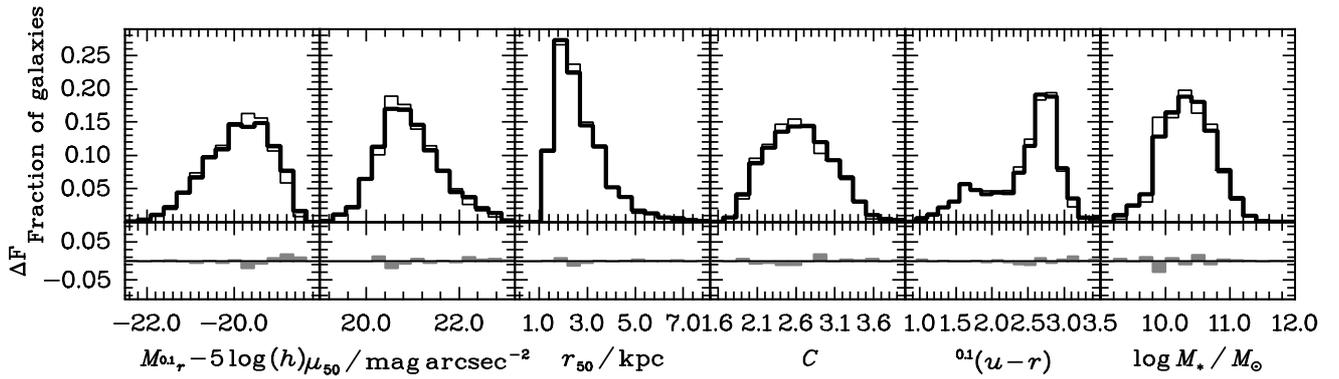}
\caption{Distributions of galaxy properties of CGs: whole sample (\textit{thick line}) and member with
spectroscopic redshift (\textit{thin line}).}
\label{fig:a1}
\end{figure*}

As explained in Sect. \ref{sec:sample}, we consider in this work a subsample of the Catalogue A
of CGs identified by \citet{McConnachie:2009}, limited to the redshift range of 
$0.06\le z \le0.18$  and apparent magnitudes $14.5\le r \le 17.77$ (2,270 galaxies).
\citet{McConnachie:2009} identified CGs following the criteria of \citet{Hickson:1982}.
In particular, the group surface brightness in the $r$-band is $\mu \le 26.0$ mag. arcsec$^{-2}$. 
Defined in this way, 1,310 galaxies ($\sim 58\%$) have spectroscopic redshift. 
Fig. \ref{fig:a1} compares the normalised distributions of galaxy parameters for all galaxies in
CGs and for those with measured spectroscopic redshift alone. No significant nor systematic differences
can be observed.

\begin{figure*}[t]
\centering
\includegraphics[width=18cm]{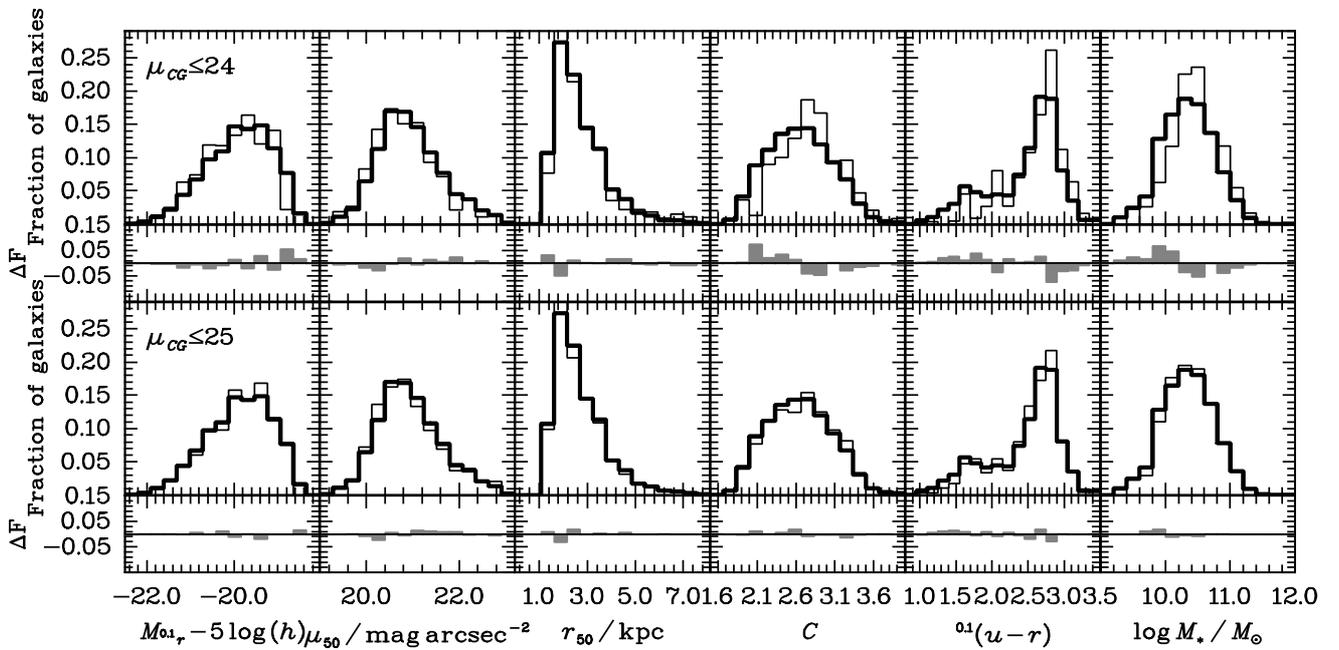}
\caption{Distributions of galaxy properties of CGs with $\mu \le 26.0$ mag. arcsec$^{-2}$ (\textit{thick line}), 
$\mu \le 25.0$ mag. arcsec$^{-2}$ and $\mu \le 24.0$ mag. arcsec$^{-2}$ (\textit{thin line}).}
\label{fig:a2}
\end{figure*}

\citet{McConnachie:2008} found that selecting CGs with a higher surface brightness threshold,
the contamination rate decreases. In \citet{McConnachie:2009} it is stated that the level 
of contamination is negligible for their Catalogue A. We checked whether our
results are affected by contamination by selecting groups with $\mu \le 25.0$ mag. arcsec$^{-2}$ 
and $\mu \le 24.0$ mag. arcsec$^{-2}$ and then comparing the properties of their galaxies with 
the members of groups with $\mu \le 26.0$ mag. arcsec$^{-2}$. 
We show the comparison between the galaxy properties of groups selected according to these
three $\mu$ values in Fig. \ref{fig:a2}.
Although galaxy properties of CGs seem similar when we consider CGs selected with 
different groups surface brightness, CGs with  $\mu \le 24.0$ mag. arcsec$^{-2}$ have a 
higher fraction of early-and red galaxies. 

\end{appendix}

\begin{acknowledgements}
We thank the anonymous referee for useful comments and suggestion which improved the paper.
This work has been supported with grants from CONICET
(PIP 11220080102603 and 11220100100336), Ministerio de Ciencia y Tecnolog\'ia 
(PID 2008/14797627), Provincia de C\'ordoba, and SECYT-UNC, Argentina.
Funding for the Sloan Digital Sky Survey (SDSS) has been provided by the 
Alfred P. Sloan 
Foundation, the Participating Institutions, the National Aeronautics and Space 
Administration, the National Science Foundation, the U.S. Department of Energy, 
the Japanese Monbukagakusho, and the Max Planck Society. The SDSS Web site is 
http://www.sdss.org/.
The SDSS is managed by the Astrophysical Research Consortium (ARC) for the 
Participating Institutions. The Participating Institutions are The University 
of Chicago, Fermilab, the Institute for Advanced Study, the Japan Participation 
Group, The Johns Hopkins University, the Korean Scientist Group, Los Alamos 
National Laboratory, the Max Planck Institut f\"ur Astronomie (MPIA), the 
Max Planck Institut f\"ur Astrophysik (MPA), New Mexico State University, 
University of Pittsburgh, University of Portsmouth, Princeton University, 
the United States Naval Observatory, and the University of Washington.
\end{acknowledgements}

\begin{thebibliography}{82}
\expandafter\ifx\csname natexlab\endcsname\relax\def\natexlab#1{#1}\fi

\bibitem[{{Abadi} {et~al.}(1999){Abadi}, {Moore}, \& {Bower}}]{Abadi:1999}
{Abadi}, M.~G., {Moore}, B., \& {Bower}, R.~G. 1999, \mnras, 308, 947

\bibitem[{{Abazajian} {et~al.}(2009){Abazajian}, {Adelman-McCarthy},
  {Ag{\"u}eros}, {Allam}, {Allende Prieto}, {An}, {Anderson}, {Anderson},
  {Annis}, {Bahcall}, \& et~al.}]{dr7}
{Abazajian}, K.~N., {Adelman-McCarthy}, J.~K., {Ag{\"u}eros}, M.~A., {et~al.}
  2009, \apjs, 182, 543

\bibitem[{{Adelman-McCarthy} {et~al.}(2008){Adelman-McCarthy}, {Ag{\"u}eros},
  {Allam}, {Allende Prieto}, {Anderson}, {Anderson}, {Annis}, {Bahcall},
  {Bailer-Jones}, {Baldry}, {Barentine}, \& et~al.}]{dr6}
{Adelman-McCarthy}, J.~K., {Ag{\"u}eros}, M.~A., {Allam}, S.~S., {et~al.} 2008,
  \apjs, 175, 297

\bibitem[{{Baldry} {et~al.}(2004){Baldry}, {Glazebrook}, {Brinkmann},
  {Ivezi{\'c}}, {Lupton}, {Nichol}, \& {Szalay}}]{Baldry:2004}
{Baldry}, I.~K., {Glazebrook}, K., {Brinkmann}, J., {et~al.} 2004, \apj, 600,
  681

\bibitem[{{Balogh} {et~al.}(2004){Balogh}, {Baldry}, {Nichol}, {Miller},
  {Bower}, \& {Glazebrook}}]{Balogh:2004}
{Balogh}, M.~L., {Baldry}, I.~K., {Nichol}, R., {et~al.} 2004, \apjl, 615, L101

\bibitem[{{Balogh} {et~al.}(2000){Balogh}, {Navarro}, \&
  {Morris}}]{Balogh:2000}
{Balogh}, M.~L., {Navarro}, J.~F., \& {Morris}, S.~L. 2000, \apj, 540, 113

\bibitem[{{Beers} {et~al.}(1990){Beers}, {Flynn}, \& {Gebhardt}}]{Beers:1990}
{Beers}, T.~C., {Flynn}, K., \& {Gebhardt}, K. 1990, \aj, 100, 32

\bibitem[{{Bernardi}(2009)}]{Bernardi:2009}
{Bernardi}, M. 2009, \mnras, 510

\bibitem[{{Bernardi} {et~al.}(2007){Bernardi}, {Hyde}, {Sheth}, {Miller}, \&
  {Nichol}}]{Bernardi:2007}
{Bernardi}, M., {Hyde}, J.~B., {Sheth}, R.~K., {Miller}, C.~J., \& {Nichol},
  R.~C. 2007, \aj, 133, 1741

\bibitem[{{Bitsakis} {et~al.}(2010){Bitsakis}, {Charmandaris}, {Le Floc'h},
  {D{\'{\i}}az-Santos}, {Slater}, {Xilouris}, \& {Haynes}}]{Bitsakis:2010}
{Bitsakis}, T., {Charmandaris}, V., {Le Floc'h}, E., {et~al.} 2010, \aap, 517,
  A75

\bibitem[{{Blanton} {et~al.}(2003){Blanton}, {Brinkmann}, {Csabai}, {Doi},
  {Eisenstein}, {Fukugita}, {Gunn}, {Hogg}, \& {Schlegel}}]{Blanton:2003}
{Blanton}, M.~R., {Brinkmann}, J., {Csabai}, I., {et~al.} 2003, \aj, 125, 2348

\bibitem[{{Blanton} {et~al.}(2005){Blanton}, {Eisenstein}, {Hogg}, {Schlegel},
  \& {Brinkmann}}]{Blanton:2005}
{Blanton}, M.~R., {Eisenstein}, D., {Hogg}, D.~W., {Schlegel}, D.~J., \&
  {Brinkmann}, J. 2005, \apj, 629, 143

\bibitem[{{Brasseur} {et~al.}(2009){Brasseur}, {McConnachie}, {Ellison}, \&
  {Patton}}]{Brasseur:2009}
{Brasseur}, C.~M., {McConnachie}, A.~W., {Ellison}, S.~L., \& {Patton}, D.~R.
  2009, \mnras, 392, 1141

\bibitem[{{Choloniewski}(1987)}]{cho87}
{Choloniewski}, J. 1987, \mnras, 226, 273

\bibitem[{{Coenda} {et~al.}(2005){Coenda}, {Donzelli}, {Muriel}, {Quintana},
  {Infante}, \& {Lambas}}]{Coenda:2005}
{Coenda}, V., {Donzelli}, C.~J., {Muriel}, H., {et~al.} 2005, \aj, 129, 1237

\bibitem[{{Coenda} \& {Muriel}(2009)}]{Coenda:2009}
{Coenda}, V. \& {Muriel}, H. 2009, \aap, 504, 347

\bibitem[{{Colless} {et~al.}(2001){Colless}, {Dalton}, {Maddox}, {Sutherland},
  {Norberg}, {Cole}, {Bland-Hawthorn}, {Bridges}, \& et~al.}]{Colless:2001}
{Colless}, M., {Dalton}, G., {Maddox}, S., {et~al.} 2001, \mnras, 328, 1039

\bibitem[{{Cortese} {et~al.}(2006){Cortese}, {Gavazzi}, {Boselli}, {Franzetti},
  {Kennicutt}, {O'Neil}, \& {Sakai}}]{Cortese:2006}
{Cortese}, L., {Gavazzi}, G., {Boselli}, A., {et~al.} 2006, \aap, 453, 847

\bibitem[{{Deng} {et~al.}(2007){Deng}, {He}, \& {Jiang}}]{Deng:2007a}
{Deng}, X.-F., {He}, J.-Z., \& {Jiang}, P. 2007, \apjl, 671, L101

\bibitem[{{Deng} {et~al.}(2008){Deng}, {He}, \& {Wu}}]{Deng:2008}
{Deng}, X.-F., {He}, J.-Z., \& {Wu}, P. 2008, \aap, 484, 355

\bibitem[{{Diaferio} {et~al.}(1994){Diaferio}, {Geller}, \&
  {Ramella}}]{Diaferio:1994}
{Diaferio}, A., {Geller}, M.~J., \& {Ramella}, M. 1994, \aj, 107, 868

\bibitem[{{D{\'{\i}}az-Gim{\'e}nez} \& {Mamon}(2010)}]{Diaz:2010}
{D{\'{\i}}az-Gim{\'e}nez}, E. \& {Mamon}, G.~A. 2010, \mnras, 409, 1227

\bibitem[{{Dressler}(1980)}]{Dressler:1980}
{Dressler}, A. 1980, \apjs, 42, 565

\bibitem[{{Eke} {et~al.}(2004){Eke}, {Baugh}, {Cole}, {Frenk}, {Norberg},
  {Peacock}, {Baldry}, {Bland-Hawthorn}, {Bridges}, {Cannon}, {Colless},
  {Collins}, {Couch}, {Dalton}, {de Propris}, {Driver}, {Efstathiou}, {Ellis},
  {Glazebrook}, {Jackson}, {Lahav}, {Lewis}, {Lumsden}, {Maddox}, {Madgwick},
  {Peterson}, {Sutherland}, \& {Taylor}}]{Eke:2004}
{Eke}, V.~R., {Baugh}, C.~M., {Cole}, S., {et~al.} 2004, \mnras, 348, 866

\bibitem[{{Gerke} {et~al.}(2007){Gerke}, {Newman}, {Faber}, {Cooper}, {Croton},
  {Davis}, {Willmer}, {Yan}, {Coil}, {Guhathakurta}, {Koo}, \&
  {Weiner}}]{Gerke:2007}
{Gerke}, B.~F., {Newman}, J.~A., {Faber}, S.~M., {et~al.} 2007, \mnras, 376,
  1425

\bibitem[{{Goto} {et~al.}(2003){Goto}, {Yamauchi}, {Fujita}, {Okamura},
  {Sekiguchi}, {Smail}, {Bernardi}, \& {Gomez}}]{Goto:2003}
{Goto}, T., {Yamauchi}, C., {Fujita}, Y., {et~al.} 2003, \mnras, 346, 601

\bibitem[{{Gunn} \& {Gott}(1972)}]{GG:1972}
{Gunn}, J.~E. \& {Gott}, J.~R.~I. 1972, \apj, 176, 1

\bibitem[{{Hansen} {et~al.}(2009){Hansen}, {Sheldon}, {Wechsler}, \&
  {Koester}}]{Hansen:2009}
{Hansen}, S.~M., {Sheldon}, E.~S., {Wechsler}, R.~H., \& {Koester}, B.~P. 2009,
  \apj, 699, 1333

\bibitem[{{Hickson}(1982)}]{Hickson:1982}
{Hickson}, P. 1982, \apj, 255, 382

\bibitem[{{Hickson} {et~al.}(1992){Hickson}, {Mendes de Oliveira}, {Huchra}, \&
  {Palumbo}}]{Hickson:1992}
{Hickson}, P., {Mendes de Oliveira}, C., {Huchra}, J.~P., \& {Palumbo}, G.~G.
  1992, \apj, 399, 353

\bibitem[{{Huchra} \& {Geller}(1982)}]{H&G:1982}
{Huchra}, J.~P. \& {Geller}, M.~J. 1982, \apj, 257, 423

\bibitem[{{Iovino} {et~al.}(2003){Iovino}, {de Carvalho}, {Gal}, {Odewahn},
  {Lopes}, {Mahabal}, \& {Djorgovski}}]{Iovino:2003}
{Iovino}, A., {de Carvalho}, R.~R., {Gal}, R.~R., {et~al.} 2003, \aj, 125, 1660

\bibitem[{{Johnson} {et~al.}(2007){Johnson}, {Hibbard}, {Gallagher},
  {Charlton}, {Hornschemeier}, {Jarrett}, \& {Reines}}]{Johnson:2007}
{Johnson}, K.~E., {Hibbard}, J.~E., {Gallagher}, S.~C., {et~al.} 2007, \aj,
  134, 1522

\bibitem[{{Kawata} \& {Mulchaey}(2008)}]{Kawata:2008}
{Kawata}, D. \& {Mulchaey}, J.~S. 2008, \apjl, 672, L103

\bibitem[{{Larson} {et~al.}(1980){Larson}, {Tinsley}, \&
  {Caldwell}}]{Larson:1980}
{Larson}, R.~B., {Tinsley}, B.~M., \& {Caldwell}, C.~N. 1980, \apj, 237, 692

\bibitem[{{Lee} {et~al.}(2004){Lee}, {Allam}, {Tucker}, {Annis}, {Johnston},
  {Scranton}, {Acebo}, {Bahcall}, {Bartelmann}, {B{\"o}hringer}, {Ellman},
  {Grebel}, {Infante}, {Loveday}, {McKay}, {Prada}, {Schneider}, {Stoughton},
  {Szalay}, {Vogeley}, {Voges}, \& {Yanny}}]{Lee:2004}
{Lee}, B.~C., {Allam}, S.~S., {Tucker}, D.~L., {et~al.} 2004, \aj, 127, 1811

\bibitem[{{Limber} \& {Mathews}(1960)}]{Limber:1960}
{Limber}, D.~N. \& {Mathews}, W.~G. 1960, \apj, 132, 286

\bibitem[{{Lynden-Bell}(1971)}]{lb71}
{Lynden-Bell}, D. 1971, \mnras, 155, 95

\bibitem[{{Maltby} {et~al.}(2010){Maltby}, {Arag{\'o}n-Salamanca}, {Gray},
  {Barden}, {H{\"a}u{\ss}ler}, {Wolf}, {Peng}, {Jahnke}, {McIntosh},
  {B{\"o}hm}, \& {van Kampen}}]{Maltby:2010}
{Maltby}, D.~T., {Arag{\'o}n-Salamanca}, A., {Gray}, M.~E., {et~al.} 2010,
  \mnras, 402, 282

\bibitem[{{Mart{\'{\i}}nez} {et~al.}(2010){Mart{\'{\i}}nez}, {Coenda}, \&
  {Muriel}}]{Martinez:2010}
{Mart{\'{\i}}nez}, H.~J., {Coenda}, V., \& {Muriel}, H. 2010, \mnras, 403, 748

\bibitem[{{Mart{\'{\i}}nez} \& {Muriel}(2006)}]{MM2:2006}
{Mart{\'{\i}}nez}, H.~J. \& {Muriel}, H. 2006, \mnras, 370, 1003

\bibitem[{{Mart{\'{\i}}nez} {et~al.}(2006){Mart{\'{\i}}nez}, {O'Mill}, \&
  {Lambas}}]{Martinez:2006}
{Mart{\'{\i}}nez}, H.~J., {O'Mill}, A.~L., \& {Lambas}, D.~G. 2006, \mnras,
  372, 253

\bibitem[{{Mart{\'{\i}}nez} \& {Zandivarez}(2012)}]{Martinez:2012}
{Mart{\'{\i}}nez}, H.~J. \& {Zandivarez}, A. 2012, \mnras, 419, L24

\bibitem[{{Mart{\'{\i}}nez} {et~al.}(2002){Mart{\'{\i}}nez}, {Zandivarez},
  {Merch{\'a}n}, \& {Dom{\'{\i}}nguez}}]{Martinez:2002}
{Mart{\'{\i}}nez}, H.~J., {Zandivarez}, A., {Merch{\'a}n}, M.~E., \&
  {Dom{\'{\i}}nguez}, M.~J.~L. 2002, \mnras, 337, 1441

\bibitem[{{McConnachie} {et~al.}(2008){McConnachie}, {Ellison}, \&
  {Patton}}]{McConnachie:2008}
{McConnachie}, A.~W., {Ellison}, S.~L., \& {Patton}, D.~R. 2008, \mnras, 387,
  1281

\bibitem[{{McConnachie} {et~al.}(2009){McConnachie}, {Patton}, {Ellison}, \&
  {Simard}}]{McConnachie:2009}
{McConnachie}, A.~W., {Patton}, D.~R., {Ellison}, S.~L., \& {Simard}, L. 2009,
  \mnras, 395, 255

\bibitem[{{McGee} {et~al.}(2009){McGee}, {Balogh}, {Bower}, {Font}, \&
  {McCarthy}}]{McGee:2009}
{McGee}, S.~L., {Balogh}, M.~L., {Bower}, R.~G., {Font}, A.~S., \& {McCarthy},
  I.~G. 2009, \mnras, 400, 937

\bibitem[{{McGee} {et~al.}(2011){McGee}, {Balogh}, {Wilman}, {Bower},
  {Mulchaey}, {Parker}, \& {Oemler}}]{McGee:2011}
{McGee}, S.~L., {Balogh}, M.~L., {Wilman}, D.~J., {et~al.} 2011, \mnras, 413,
  996

\bibitem[{{Mendel} {et~al.}(2011){Mendel}, {Ellison}, {Simard}, {Patton}, \&
  {McConnachie}}]{Mendel:2011}
{Mendel}, J.~T., {Ellison}, S.~L., {Simard}, L., {Patton}, D.~R., \&
  {McConnachie}, A.~W. 2011, ArXiv e-prints

\bibitem[{{Mendes de Oliveira} \& {Hickson}(1994)}]{MOH:1994}
{Mendes de Oliveira}, C. \& {Hickson}, P. 1994, \apj, 427, 684

\bibitem[{{Merch{\'a}n} \& {Zandivarez}(2002)}]{Merchan:2002}
{Merch{\'a}n}, M. \& {Zandivarez}, A. 2002, \mnras, 335, 216

\bibitem[{{Merch{\'a}n} \& {Zandivarez}(2005)}]{Merchan&Zandivarez:2005}
{Merch{\'a}n}, M.~E. \& {Zandivarez}, A. 2005, \apj, 630, 759

\bibitem[{{Moore} {et~al.}(1993){Moore}, {Frenk}, \& {White}}]{Moore:1993}
{Moore}, B., {Frenk}, C.~S., \& {White}, S.~D.~M. 1993, \mnras, 261, 827

\bibitem[{{Moore} {et~al.}(1998){Moore}, {Lake}, \& {Katz}}]{Moore:1998}
{Moore}, B., {Lake}, G., \& {Katz}, N. 1998, \apj, 495, 139

\bibitem[{{Nair} {et~al.}(2010){Nair}, {van den Bergh}, \&
  {Abraham}}]{Nair:2010}
{Nair}, P.~B., {van den Bergh}, S., \& {Abraham}, R.~G. 2010, \apj, 715, 606

\bibitem[{{Oemler}(1974)}]{Oemler:1974}
{Oemler}, A.~J. 1974, \apj, 194, 1

\bibitem[{{Petrosian}(1976)}]{petro76}
{Petrosian}, V. 1976, \apjl, 209, L1

\bibitem[{{Rasmussen} {et~al.}(2008){Rasmussen}, {Ponman}, {Verdes-Montenegro},
  {Yun}, \& {Borthakur}}]{Rasmussen:2008}
{Rasmussen}, J., {Ponman}, T.~J., {Verdes-Montenegro}, L., {Yun}, M.~S., \&
  {Borthakur}, S. 2008, \mnras, 388, 1245

\bibitem[{{Sandage} {et~al.}(1979){Sandage}, {Tammann}, \& {Yahil}}]{sty}
{Sandage}, A., {Tammann}, G.~A., \& {Yahil}, A. 1979, \apj, 232, 352

\bibitem[{{Sandage} \& {Visvanathan}(1978{\natexlab{a}})}]{SV:1978b}
{Sandage}, A. \& {Visvanathan}, N. 1978{\natexlab{a}}, \apj, 225, 742

\bibitem[{{Sandage} \& {Visvanathan}(1978{\natexlab{b}})}]{SV:1978a}
{Sandage}, A. \& {Visvanathan}, N. 1978{\natexlab{b}}, \apj, 223, 707

\bibitem[{{Schechter}(1976)}]{Schechter:1976}
{Schechter}, P. 1976, \apj, 203, 297

\bibitem[{{Schlegel} {et~al.}(1998){Schlegel}, {Finkbeiner}, \&
  {Davis}}]{sch98}
{Schlegel}, D.~J., {Finkbeiner}, D.~P., \& {Davis}, M. 1998, \apj, 500, 525

\bibitem[{{Schmidt}(1968)}]{Schmidt:68}
{Schmidt}, M. 1968, \apj, 151, 393

\bibitem[{{Shen} {et~al.}(2003){Shen}, {Mo}, {White}, {Blanton}, {Kauffmann},
  {Voges}, {Brinkmann}, \& {Csabai}}]{Shen:2003}
{Shen}, S., {Mo}, H.~J., {White}, S.~D.~M., {et~al.} 2003, \mnras, 343, 978

\bibitem[{{Springel} {et~al.}(2005){Springel}, {White}, {Jenkins}, {Frenk},
  {Yoshida}, {Gao}, {Navarro}, {Thacker}, {Croton}, {Helly}, {Peacock}, {Cole},
  {Thomas}, {Couchman}, {Evrard}, {Colberg}, \& {Pearce}}]{Springel:2005}
{Springel}, V., {White}, S.~D.~M., {Jenkins}, A., {et~al.} 2005, \nat, 435, 629

\bibitem[{{Strateva} {et~al.}(2001){Strateva}, {Ivezi{\'c}}, {Knapp},
  {Narayanan}, {Strauss}, {Gunn}, {Lupton}, {Schlegel}, {Bahcall}, \&
  et~al.}]{Strateva:2001}
{Strateva}, I., {Ivezi{\'c}}, {\v Z}., {Knapp}, G.~R., {et~al.} 2001, \aj, 122,
  1861

\bibitem[{{Strauss} {et~al.}(2002){Strauss}, {Weinberg}, {Lupton}, {Narayanan},
  {Annis}, {Bernardi}, {Blanton}, {Burles}, {Connolly}, {Dalcanton}, {Doi},
  {Eisenstein}, \& et~al.}]{Strauss:2002}
{Strauss}, M.~A., {Weinberg}, D.~H., {Lupton}, R.~H., {et~al.} 2002, \aj, 124,
  1810

\bibitem[{{Taylor} {et~al.}(2011){Taylor}, {Hopkins}, {Baldry}, {Brown},
  {Driver}, {Kelvin}, {Hill}, {Robotham}, {Bland-Hawthorn}, {Jones}, {Sharp},
  {Thomas}, {Liske}, {Loveday}, {Norberg}, {Peacock}, {Bamford}, {Brough},
  {Colless}, {Cameron}, {Conselice}, {Croom}, {Frenk}, {Gunawardhana},
  {Kuijken}, {Nichol}, {Parkinson}, {Phillipps}, {Pimbblet}, {Popescu},
  {Prescott}, {Sutherland}, {Tuffs}, {van Kampen}, \&
  {Wijesinghe}}]{Taylor:2011}
{Taylor}, E.~N., {Hopkins}, A.~M., {Baldry}, I.~K., {et~al.} 2011, \mnras, 418,
  1587

\bibitem[{{Toomre} \& {Toomre}(1972)}]{TT:1972}
{Toomre}, A. \& {Toomre}, J. 1972, \apj, 178, 623

\bibitem[{{Torres-Flores} {et~al.}(2009){Torres-Flores}, {Mendes de Oliveira},
  {de Mello}, {Amram}, {Plana}, {Epinat}, \&
  {Iglesias-P{\'a}ramo}}]{Torres:2009}
{Torres-Flores}, S., {Mendes de Oliveira}, C., {de Mello}, D.~F., {et~al.}
  2009, \aap, 507, 723

\bibitem[{{Tovmassian} {et~al.}(2006){Tovmassian}, {Plionis}, \&
  {Torres-Papaqui}}]{Tovmassian:2006}
{Tovmassian}, H., {Plionis}, M., \& {Torres-Papaqui}, J.~P. 2006, \aap, 456,
  839

\bibitem[{{Tzanavaris} {et~al.}(2010){Tzanavaris}, {Hornschemeier},
  {Gallagher}, {Johnson}, {Gronwall}, {Immler}, {Reines}, {Hoversten}, \&
  {Charlton}}]{Tzanavaris:2010}
{Tzanavaris}, P., {Hornschemeier}, A.~E., {Gallagher}, S.~C., {et~al.} 2010,
  \apj, 716, 556

\bibitem[{{von der Linden} {et~al.}(2007){von der Linden}, {Best}, {Kauffmann},
  \& {White}}]{vonderlinden:2007}
{von der Linden}, A., {Best}, P.~N., {Kauffmann}, G., \& {White}, S.~D.~M.
  2007, \mnras, 379, 867

\bibitem[{{Walker} {et~al.}(2010){Walker}, {Johnson}, {Gallagher}, {Hibbard},
  {Hornschemeier}, {Tzanavaris}, {Charlton}, \& {Jarrett}}]{Walker:2010}
{Walker}, L.~M., {Johnson}, K.~E., {Gallagher}, S.~C., {et~al.} 2010, \aj, 140,
  1254

\bibitem[{{Weinmann} {et~al.}(2009){Weinmann}, {Kauffmann}, {van den Bosch},
  {Pasquali}, {McIntosh}, {Mo}, {Yang}, \& {Guo}}]{Weinmann:2009}
{Weinmann}, S.~M., {Kauffmann}, G., {van den Bosch}, F.~C., {et~al.} 2009,
  \mnras, 394, 1213

\bibitem[{{Weinmann} {et~al.}(2006){Weinmann}, {van den Bosch}, {Yang}, \&
  {Mo}}]{Weinmann:2006}
{Weinmann}, S.~M., {van den Bosch}, F.~C., {Yang}, X., \& {Mo}, H.~J. 2006,
  \mnras, 366, 2

\bibitem[{{Wetzel} {et~al.}(2011){Wetzel}, {Tinker}, \& {Conroy}}]{Wetzel:2011}
{Wetzel}, A.~R., {Tinker}, J.~L., \& {Conroy}, C. 2011, ArXiv e-prints

\bibitem[{{Wilman} {et~al.}(2009){Wilman}, {Oemler}, {Mulchaey}, {McGee},
  {Balogh}, \& {Bower}}]{Wilman:2009}
{Wilman}, D.~J., {Oemler}, Jr., A., {Mulchaey}, J.~S., {et~al.} 2009, \apj,
  692, 298

\bibitem[{{Yang} {et~al.}(2007){Yang}, {Mo}, {van den Bosch}, {Pasquali}, {Li},
  \& {Barden}}]{Yang:2007}
{Yang}, X., {Mo}, H.~J., {van den Bosch}, F.~C., {et~al.} 2007, \apj, 671, 153

\bibitem[{{York} {et~al.}(2000){York}, {Anderson}, {Anderson}, {Annis},
  {Bahcall}, \& et~al.}]{York:2000}
{York}, D.~G., {Anderson}, Jr., J.~E., {Anderson}, S.~F., {et~al.} 2000, \aj,
  120, 1579

\bibitem[{{Zandivarez} \& {Mart{\'{\i}}nez}(2011)}]{ZM11}
{Zandivarez}, A. \& {Mart{\'{\i}}nez}, H.~J. 2011, \mnras, 415, 2553

\end{thebibliography}

\end{document}